\documentclass{mn2e}
\usepackage[dvips]{graphicx}

\begin{document}

  \title[Effects of feedback on the formation of disks]{Effects
    of Supernova Feedback on the Formation of Galaxy Disks}

\author[Scannapieco et al.]{Cecilia Scannapieco $^{1}$\thanks{E-mail: cecilia@mpa-garching.mpg.de},
Patricia B. Tissera$^{2,3}$, Simon D.M. White$^{1}$ and Volker Springel$^{1}$ \\
$^1$ Max-Planck Institute for Astrophysics, Karl-Schwarzchild Str. 1, D85748, Garching, Germany.\\
$^2$ Instituto de Astronom\'{\i}a y F\'{\i}sica del Espacio, Casilla de Correos 67,
Suc. 28, 1428, Buenos Aires, Argentina.\\
$^3$  Consejo Nacional de Investigaciones Cient\'{\i}ficas
y T\'ecnicas, CONICET, Argentina.\\ 
}

   \maketitle

   \begin{abstract}
We use cosmological simulations in order to study the effects
of supernova (SN) feedback on the  formation of a 
Milky Way-type galaxy of virial mass  $\sim 10^{12}\
h^{-1}$M$_\odot$. We analyse 
a set of simulations
run with the code described by Scannapieco et al. (2005, 2006), 
where we have tested our star formation and feedback
prescription using isolated galaxy models.  Here we extend this 
work by simulating the formation of a
galaxy  in its proper cosmological framework,
focusing on the ability of the model to form a disk-like structure
in rotational support.
We find that SN  feedback plays a fundamental role
in the evolution of the simulated galaxy, efficiently regulating the star
formation activity, pressurizing the gas and generating
mass-loaded galactic winds. These processes affect several
galactic properties such as final stellar mass, morphology,
angular momentum, chemical properties, and final gas and baryon
fractions.
In particular, we find that our model is able to reproduce
extended disk components with high specific angular momentum
and a significant fraction of young stars.
The galaxies are also
found to have significant spheroids composed almost entirely of
stars formed at early times.
We find that most combinations of the input
parameters yield  disk-like
components, although with different sizes and thicknesses, 
 indicating that the code
can  form disks without fine-tuning the implemented physics. 
We also show how our model scales to smaller systems. By analysing simulations of virial masses
$10^{9}\ h^{-1}$M$_\odot$ and $10^{10}\ h^{-1}$M$_\odot$, we find that
the smaller the galaxy, the stronger the SN feedback effects.

   \end{abstract}


\section{Introduction}

The formation of spiral galaxies is currently a subject
of considerable interest in cosmology. 
From a theoretical point of view, 
the formation of disk configurations is
well understood as a result of specific angular momentum conservation 
during the collapse and relaxation of  gas within 
dark matter potential wells (Fall \& Efstathiou 1980; Mo,
Mao \& White 1998).
Dark matter and baryons aquire similar specific angular
momenta from tidal torques before collapse (Hoyle 1953; Peebles 1969; White 1984). Provided the gas
conserves this angular momentum, its collapse leads to a disk-like
configuration
at the centre of its dark matter halo. If stars  form
from this rotationally supported gas, then spiral disks like those observed can develop.
However, centrifugally supported stellar disks are thickened or
destroyed
 by rapid fluctuations in
the gravitational potential such as those produced by the
accretion of satellite galaxies (Toth \& Ostriker 1992; Quinn, Hernquist \& Fullagar 1993;
Velazquez \& White 1999). 
The presence of thin disks in many present-day spirals  
suggests that  they had a relatively quiet mass accretion history,
at least in the recent past. 

Numerical simulations have become an important tool for 
studying galaxy formation due to their
ability to describe the nonlinear, coupled  evolution of dark matter and
baryons in its cosmological context.
However, this has turned out to be an extremely challenging problem
because of the need to describe simultaneously the large-scale
dynamics of galaxy assembly and the small-scale processes
related to star formation and evolution. State-of-the-art simulations
have implemented schematic recipes for  radiative cooling,
star formation, feedback from supernovae (SN), stellar winds,
black hole formation and active galactic nuclei.
Although these are successful in many respects,
they have had difficulty 
reproducing extended disk-dominated galaxies in typical
dark matter haloes.
The main problem is  so-called
 {\it catastrophic angular momentum loss}, 
which arises  when the baryonic components transfer most of 
their angular momentum
to the dark matter  during
condensation into the central regions,
forming  dominant spheroidal stellar systems
and  disks that
are too small  compared to observations
(e.g. Navarro \& Benz 1991; Navarro \& White 1994).

The formation of disks is heavily influenced by details
of the gas cooling process.  Disks that are less centrally concentrated
and with higher specific angular momentum are
obtained if cooling is artificially suppressed
during the assembly of dark matter haloes (Weil, Eke \&
Efstathiou 1998; Eke, Efstathiou \& Wright 2000).
Many recent studies have attempted
to prevent
the early cooling of gas and the loss of angular momentum,
some of them successfully  reproducing disks
as big as observed spirals
(e.g. Abadi et al. 2003; Governato et al. 2004;
Robertson et al. 2004; Okamoto et al. 2005; Governato et al. 2007).
These studies have pointed out the importance of stellar
feedback as a key process providing pressure support
to the gas and preventing its rapid collapse within the haloes,
thereby regulating  star formation activity.
However, it seems that  fine-tuning of the models is required in order
to reproduce the observed properties of disk galaxies in
a hierarchical scenario, and a detailed understanding of disk formation
remains elusive.
Insufficient numerical resolution or a poor description of the
interstellar medium (which influences the efficiency of the
stellar feedback) have  also been proposed as causes
for the inability to reproduce convincing disk galaxies in simulations
(e.g. Okamoto et al. 2003; Governato et al. 2004).

Over the years, many authors 
have included star formation and feedback recipes  both
in mesh-based and in SPH codes
in order to study
the effects of SN feedback
on galaxy properties such
as star formation histories, metallicity evolution,
generation of galactic outflows, abundance of satellites
and galaxy scaling relations
(e.g. Katz \& Gunn 1991; Cen \& Ostriker 1992;
Navarro \& White 1993; Metzler \& Evrard 1994; Steinmetz \& M\"uller 1995; 
Gerritsen \& Icke 1997; Yepes et al. 1997;  Cen \& Ostriker 1999;
Sommer-Larsen, Gelato \& Vedel 1999; Thacker \& Couchman 2000; Kay et
al. 2002;  Semelin \& Combes 2002; Abadi et al. 2003;
Marri \& White 2003;  Sommer-Larsen, G\"otz \& Portinari 2003;
Springel \&
Hernquist 2003; Brook et al. 2004; Okamoto et al. 2005; Oppenheimer \& Dav\'e 2006;
Stinson et al. 2006; Governato et al. 2007;
Dalla Vecchia \& Schaye 2008; Dubois \& Teyssier 2008; Finlator \& Dav\'e 2008).
All these feedback implementations require arbitrary  and {\it ad hoc} 
elements because the relevant dynamical processes cannot be
resolved by the simulations and so have to be introduced as
``sub-grid'' models. As a result, their validity is uncertain and the
conclusions to be drawn from their successes and failures are
correspondingly
unclear. A further difficulty is introduced by
numerical limitations which have so far prevented the simulation 
at high resolution of a large and representative sample of
galaxies in their proper cosmological setting; typically, the
evolution of a small number of selected objects is studied in detail,
often chosen as those most likely to produce disk galaxies. It is
thus uncertain whether existing models can successfully reproduce
the variety of disk galaxies observed in the Universe. In particular,
no $\Lambda$CDM simulation has yet managed to produce a late-type
spiral like M33 or M101.

In Scannapieco et al. (2005, 2006), we have presented a new
model for the effects of SN feedback in cosmological simulations.
This model was implemented within the Tree-PM SPH code 
{\small GADGET-2} (Springel
\& Hernquist 2002; Springel 2005), and is described in detail in
 Scannapieco et al. (2005, 2006).
Our model includes a scheme to treat a multiphase interstellar
medium which allows the representation of a co-spatial mixture of
cold and hot interstellar medium (ISM) components without introducing scale-dependent
parameters. Energy and chemical feedback by Type II and
Type Ia SNe, as well as metal-dependent gas cooling
(Sutherland \& Dopita 1993), are also considered.
The model is intended to be as simple
as possible and to introduce the fewest  input parameters possible.
It was specially designed 
to avoid the inclusion of mass-scale dependent parameters.
This makes it suitable for running cosmological simulations
where, at any given time,  systems of different masses
are forming simultaneously.
In Scannapieco et al. (2005, 2006) we tested the code
by studying the formation of disk galaxies of different
mass from idealized initial conditions.
Our scheme can reproduce several important effects
in galaxy evolution: substantial metal-enhanced galactic winds are
produced, redistributing mass and metals and generating
a self-regulated cycle for star formation.
The galactic winds are generated naturally in our model 
when the multiphase gas model and SN feedback are both included.
This is an important result which allows us to consistently
study the formation of galaxies together with the chemical
enrichment  and the redistribution of matter within
the galaxies. Potentially it should also allow investigation
of other aspects such as the enrichment of the intergalactic medium.
Our model reproduces the expected dependence on
galaxy mass:
while star formation is
suppressed at most by a factor of a few in massive galaxies,  in low-mass
systems the effects can be much larger, giving star formation an episodic,
bursty character.
In this paper,
we extend the work of Scannapieco et al. (2005, 2006) 
by studying the formation of disk galaxies
in their proper cosmological context. This allows a consistent
treatment of processes such as mergers, interactions,
inflows, tidal stripping, etc. We thus carry out
higher resolution simulations from realistic initial conditions
and follow the hierarchical growth of a galaxy similar
in size and spin to our Milky Way.
We focus on the effects of SN feedback on
star formation and  on   galaxy
morphology,  emphasizing the dependence of the predictions on input parameters.

Our paper is organized as follows.
In Section~\ref{sims} we 
describe the simulation code and list the characteristics of
our simulations, briefly describing the
inputs  which are
relevant for this work.
In Section~\ref{results} we explore
the effects of SN feedback on
star formation  (subsection~\ref{sect_sfr}),  
galaxy morphology (subection~\ref{sect_morph})
 and evolution of specific angular momentum
 (subsection~\ref{sect_j_age_fe}).
We also study the formation process of the disk
components found in our simulated galaxies
(subsection~\ref{stell_age}) 
and the resulting chemical properties (subsection~\ref{chem_prop}). 
In Section~\ref{dwarf} we investigate how SN feedback proceeds
for smaller galaxies. Finally, we give our conclusions in Section~\ref{conclu}.

\section{The simulation code}
\label{sims}

The simulations in this paper were run with
the code described in Scannapieco et al.
(2005, 2006), which is an extension of
the parallel Tree-PM SPH code {\small GADGET-2} 
(Springel \& Hernquist 2002; Springel 2005).
The code includes star formation, metal-dependent
cooling, chemical enrichment and energy feedback
by Type II and Type Ia Supernovae, in addition to a
multiphase treatment for the gas component.
At the end of this Section, we describe those input
parameters of our model that are particularly relevant
for the present study, and
we refer the interested reader to
Scannapieco et al. (2005) for details on
the implementation of star formation,
chemical enrichment and metal-dependent cooling,
and to Scannapieco et al. (2006) for the
description of the multiphase gas 
and energy feedback models.
Note that our multiphase treatment
and our star formation and feedback models 
differ substantially from those of Springel
\& Hernquist (2003) although we  do adopt their
treatment of the UV background.

We focus on a study of the formation of disk galaxies
in their cosmological context. For this purpose we 
simulate a system with $z=0$ halo mass  $\sim 10^{12}$ $h^{-1}$ M$_\odot$,
extracted from a large cosmological simulation and
resimulated with improved resolution.
Our simulation  runs from $z=38$ to $z=0$ 
and adopts a $\Lambda$CDM Universe
with the following cosmological parameters: 
$\Omega_\Lambda=0.7$, $\Omega_{\rm m}=0.3$,
$\Omega_{\rm b}=0.04$,  a normalization of the power spectrum of
$\sigma_8=0.9$ and 
$H_0=100\ h$ km s$^{-1}$ Mpc$^{-1}$ with $h=0.7$.
The particle mass is
$1.6\times 10^7$ for dark matter and $2.4\times 10^6$ $h^{-1}$
M$_\odot$  for baryonic particles, and we use  a maximum gravitational 
softening of $0.8\ h^{-1}$ kpc
for gas, dark matter and star particles.

At $z=0$ the halo of our galaxy has spin parameter of $\lambda=0.03$
and 
it contains $\sim 1.2\times 10^5$ dark matter
and $\sim 1.5\times 10^5$  baryonic particles within the virial radius.  
It was selected to have no major mergers
since $z=1$ in order to give time for a disk to form.
Because the selected halo is similar in mass and spin to the Milky
Way and has a relatively quiet  merger history in the recent
past, we thus expect this halo to host a disk galaxy. However,
given our incomplete knowledge of galaxy formation
processes, this expectation is not necessarily fully correct.

\begin{table*} 
\begin{small}
\caption{Principal characteristics of our simulations: fraction of energy
(and metals) distributed into the cold phase ($\epsilon_c$), energy
per SN ($E_{\rm SN}$ in units of $10^{51}$ ergs) and star
formation efficiency ($c$). We also list  dark matter, gas and stellar
mass in units of $10^{10}\ h^{-1}$M$_\odot$, gas fraction and baryonic
fraction 
both within the virial radius (superscripts
'$200$') and within twice the optical radius (superscripts '$2r_{\rm
opt}$'). We show the optical radius for the different
simulations, in units of $h^{-1}$kpc.
The  fraction of stellar mass  formed at $z<1$ is also shown. All
quantities are computed at $z=0$.}
\vspace{0.1cm}
\label{simulations_table}
\begin{center}
\begin{tabular}{lccccccccccccccccc}
\hline
Test  & $\epsilon_c$&  $E_{\rm SN}$ & $c$  & $M_{\rm DM}^{200}$ &
$M_{\rm gas}^{200}$ & $M_{\rm star}^{200}$  & $f_{\rm g}^{200}$ &
$f_{\rm bar}^{200}$
& $r_{\rm opt}$ & $M_{\rm DM}^{2\rm{r_{\rm opt}}}$ & $M_{\rm gas}^{2\rm{r_{\rm opt}}}$ & $M_{\rm star}^{2\rm{r_{\rm opt}}}$  & $f_{\rm g}^{2\rm{r_{\rm opt}}}$  & $f_{\rm bar}^{2\rm{r_{\rm opt}}}$  &  $f_*^{\rm z<1}$\\\hline

NF       & -   & -   & 0.1  & 196.4 & 6.9 & 24.6 & 0.22 & 0.14 & 12.8 & 34.8 & 0.74 & 14.9 & 0.05 & 0.31 & 0.11 \\
F-0.3    & 0.3 &   1 & 0.1  & 188.1 & 5.2 &  5.6 & 0.48 & 0.05 & 12.5 & 30.2 & 0.51 &  4.4 & 0.11 & 0.14 & 0.16\\
F-0.5    & 0.5 &   1 & 0.1  & 190.4 & 6.2 &  6.6 & 0.49 & 0.06 & 12.0 & 29.6 & 0.37 &  5.5 & 0.06 & 0.17 & 0.25 \\
F-0.9    & 0.9 &   1 & 0.1  & 196.2 & 8.2 &  7.7 & 0.52 & 0.08 & 15.1 & 38.9 & 0.77 &  6.6 & 0.10 & 0.16 & 0.30 \\
E-0.3    & 0.5 & 0.3 & 0.1  & 196.4 & 6.8 & 16.8 & 0.29 & 0.11 & 11.0 & 30.3 & 0.53 & 12.9 & 0.04 & 0.31 & 0.24 \\
E-0.7    & 0.5 & 0.7 & 0.1  & 192.9 & 6.1 &  9.1 & 0.40 & 0.07 & 10.0 & 25.6 & 0.54 &  7.4 & 0.07 & 0.24 & 0.27 \\
E-3      & 0.5 &   3 & 0.1  & 187.2 & 5.6 &  1.7 & 0.77 & 0.04 & 13.6 & 31.8 & 0.27 &  1.2 & 0.18 & 0.05 & 0.24 \\
C-0.01   & 0.5 &   1 & 0.01 & 184.1 & 4.0 & 11.9 & 0.25 & 0.08 &  8.9 & 21.5 & 0.12 &  9.8 & 0.01 & 0.32 & 0.21 \\
C-0.5    & 0.5 &   1 & 0.5  & 193.1 & 7.0 &  5.6 & 0.56 & 0.06 & 14.5 & 37.3 & 0.49 &  4.4 & 0.10 & 0.12 & 0.31 \\\hline
\end{tabular}
\end{center}
\end{small}
\end{table*}

We have run a series of simulations in order to study the dependence
of the results on assumed input parameters and to see if the code can
produce disk systems. Because of this focus, we decided to vary only
the parameters related to star formation and energy feedback, and we
fixed those related to chemical enrichment and phase decoupling.  In
all our simulations we have assumed a SNIa rate of $0.3$ relative to
SNII and a SNIa lifetime uniformally distributed in the range
$[0.1,1]$ Gyr.  We have adopted the instantaneous recycling
approximation for SNII, we have used metal-dependent cooling
(Sutherland \& Dopita 1993) and we have adopted a Salpeter Initial
Mass Function.   Chemical yields from Woosley \& Weaver (1995) and
Thielemann et al. (1993) have been used for SNII and SNIa,
respectively.  The multiphase gas model of Scannapieco et al. (2006)
has been turned on for all experiments. This model {\it decouples}
particles with dissimilar properties, preventing them from being SPH
neighbours. In all the simulations presented in this paper, we have
used $\alpha=50$ for the decoupling parameter which sets the
difference in entropy required for particles to be decoupled (see
Scannapieco et al. 2006 for details).

For the star formation and feedback models, there are
three relevant input parameters:
a) the star formation efficiency $c$, b) the
energy $E_{\rm SN}$ released per SN\footnote{
In Scannapieco et al. (2006) we used a parameter
$\epsilon_{\rm r}$ which denotes the fraction of
the SN energy assumed to be lost by radiation. In practice,
this is equivalent to assume a lower value for $E_{\rm SN}$.
In this work,  we decided to use $E_{\rm SN}$ instead
of  $\epsilon_{\rm r}$.} 
and c) the
so-called {\it feedback parameter} $\epsilon_c$.
In our model, we assume that gas particles
are eligible for star formation if
they are denser than a critical value ($\rho>\rho_{crit} = 
7\times 10^{-26} {\rm g}\  {\rm cm}^{-3}$ where $\rho$ denotes gas density) 
and lie in a convergent
flow. For these particles, we assume a star
formation rate (SFR) per unit volume equal to
\begin{equation}
\dot\rho_\star = c\,\frac{\rho}{\tau_{\rm dyn}},
\end{equation}
where $c$ is the star formation efficiency, $\rho_*$ denotes
stellar density,   and $\tau_{\rm dyn} = 1 / \sqrt{4\pi G\rho}$
is the local dynamical time of the particle.
The star formation efficiency then sets the efficiency
at which stars are formed or, equivalently, the typical
time-scale of the star formation process.
Once star particles are formed, they are assumed
to give rise to both SNII and SNIa explosions
(see Scannapieco et al. 2005 for details on
the implementation of star formation and of SN explosions).
Each SN event will release an amount of energy,
given by the number of exploding stars within the stellar
particle multiplied by the energy per SN, 
 $E_{\rm SN}$. Finally,  a feedback parameter is introduced in our
model when implementing the transfer of SN
energy to the gas. This is based on a separation
of gas particles into two phases called
{\it hot} and {\it cold} (split by  assumed thresholds in 
density and temperature). When SN explosions take place,
the energy is distributed  as thermal feedback in
a fixed proportion to these
two phases, the cold phase receives a fraction 
$\epsilon_c$  and the remaining fraction
$1-\epsilon_c$
is distributed to the hot neighbours.
While hot particles receive the energy instantaneously,
each cold particle instead accumulates the SN energy
until it is large enough to raise its entropy
to the level of its local hot neighbours.
At each time we define for each cold particle a local
hot phase and  compare the average entropy of the identified hot neighbours
with the entropy that the cold particle would have if it
absorbed the accumulated energy.
This gives the code the flexibility to accomodate
itself to local conditions,
since different  systems will have different
typical hot-phase entropies and these will change with time.
The parameter $\epsilon_c$
is therefore related to the mode of feedback 
in a non-trivial way
(Scannapieco et al. 2006).
Together with the values adopted
for the star formation efficiency
and energy released per SN,  the feedback parameter
determines the details of the evolution
of the galaxy.

A list of the simulations we have carried out
is shown in Table \ref{simulations_table}.
They are divided into three series: F, E and C refering
to the parameter which is varied in each series. Simulations
F were run with different feedback parameter ($\epsilon_c$), simulations E with
different energy per SN ($E_{\rm SN}$) and simulations C with different
star formation efficiency ($c$). 
Note that  F-0.5 is a standard case which belongs to all
three groups of simulations.
We have also run a simulation without
 energy feedback  (although including chemical
enrichment,
metal-dependent cooling and the multiphase gas treatment) in order to
highlight
the effects of this process. This simulation is referred to as NF.

\begin{figure}
\includegraphics[width=80mm]{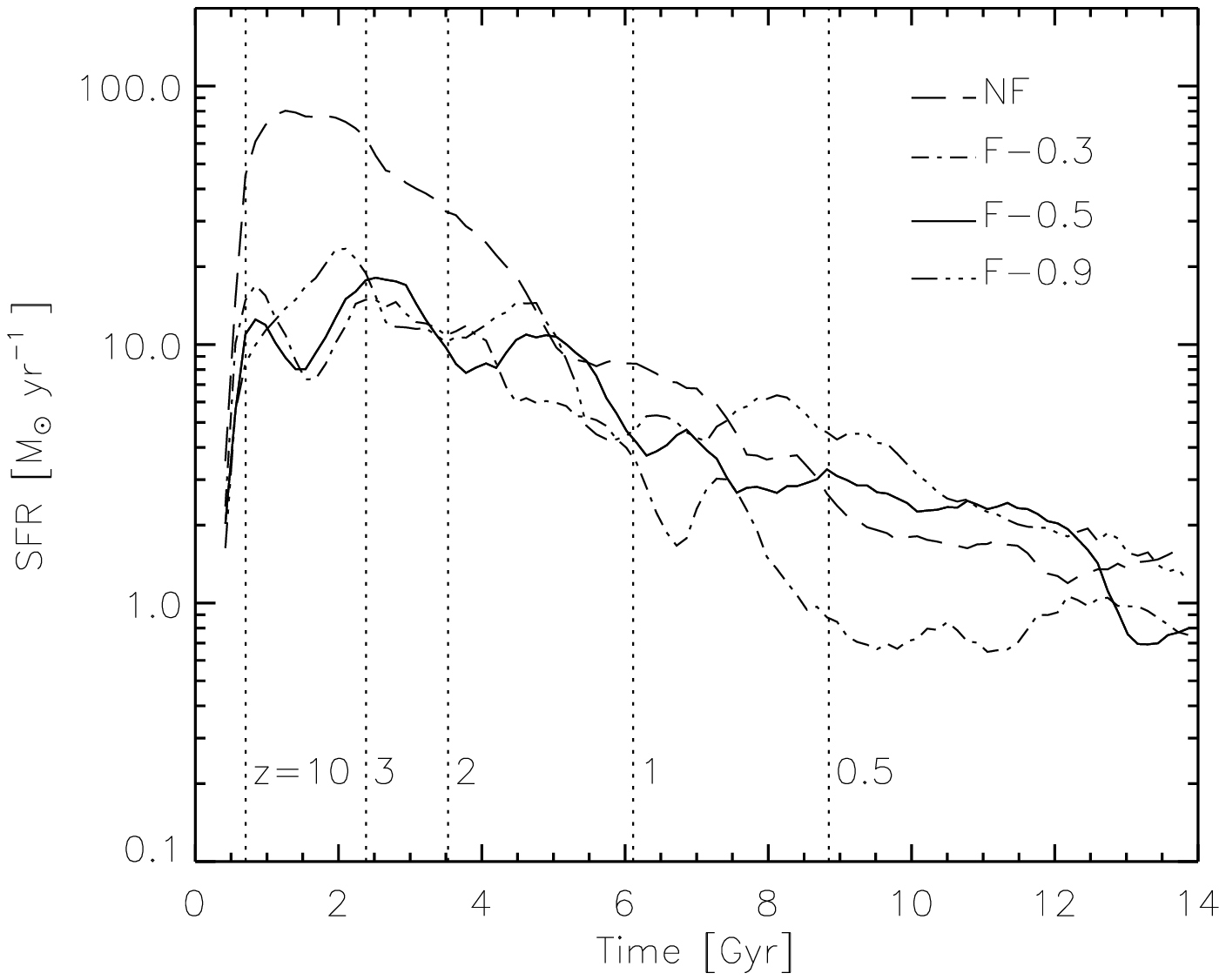}
\includegraphics[width=80mm]{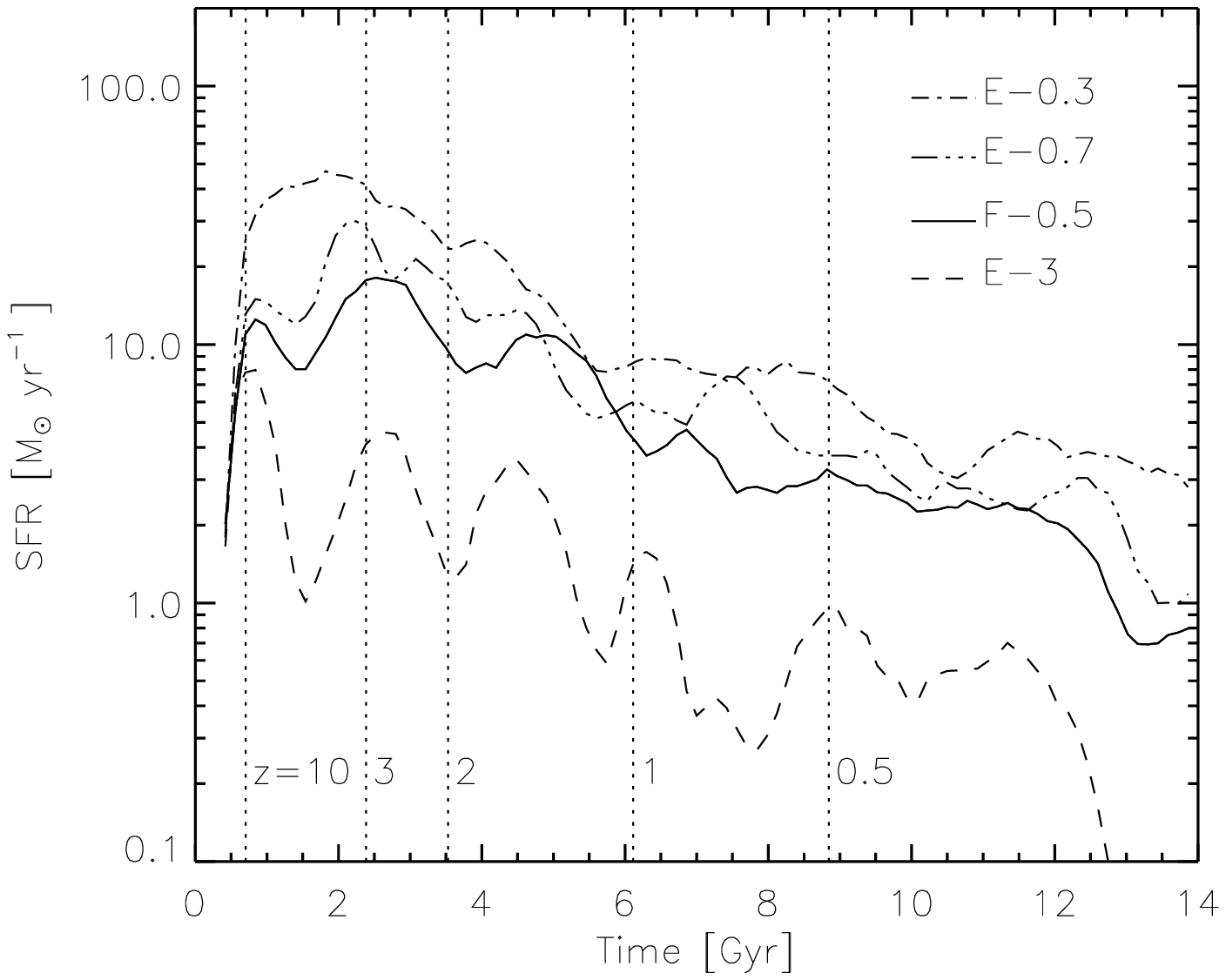}
\includegraphics[width=80mm]{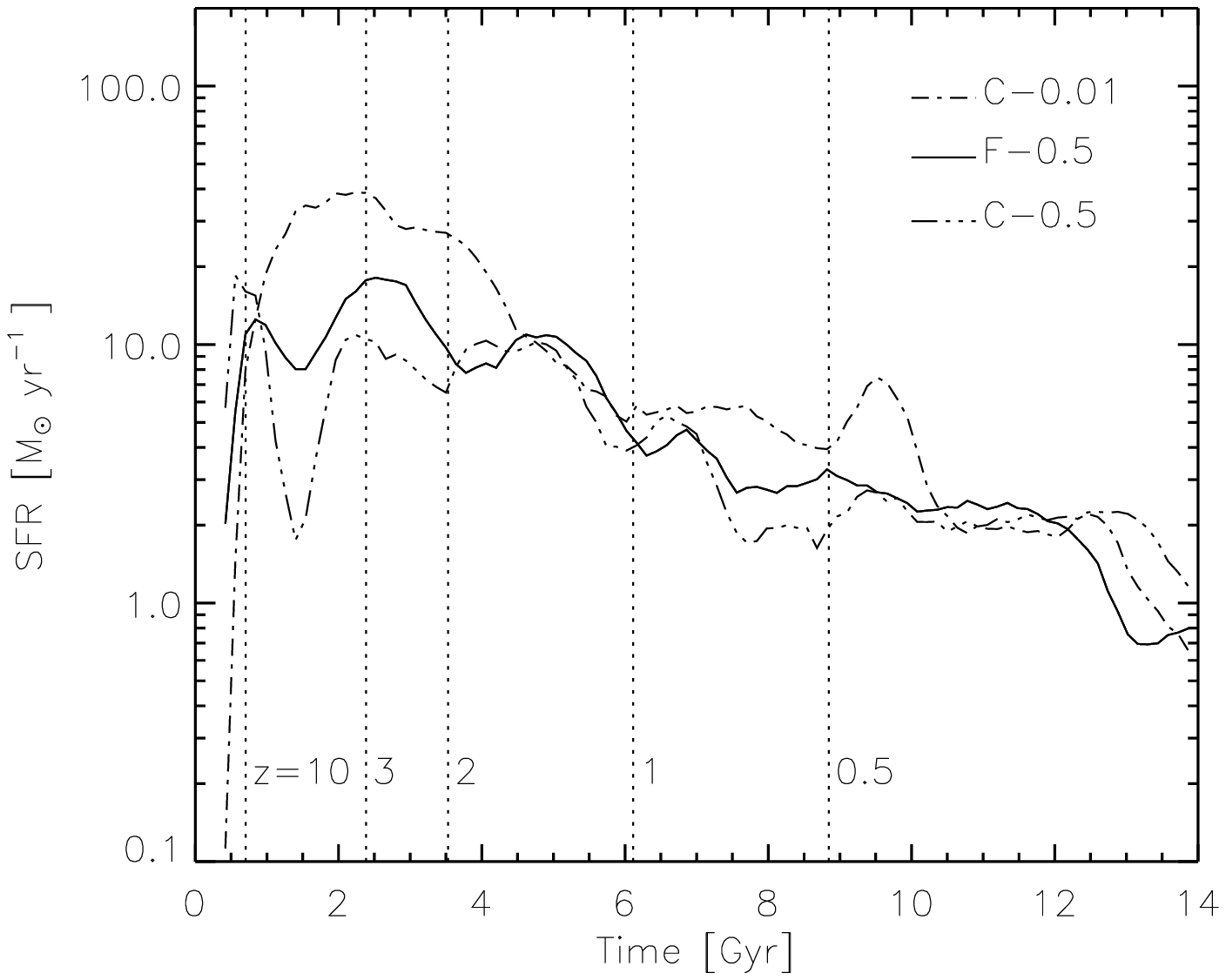}
\caption{Star formation rates for the three different series of
experiments. 
Upper panel: results for varying feedback parameter (F-0.3, F-0.5 and
F-0.9),
as well as for the no-feedback simulation (NF).
Middle panel: results for varying energy per SN (E-0.3,
E-0.7, F-0.5 ($E_{\rm SN}=1\times 10^{51}$ ergs) and E-3). Lower panel:
results for varying star formation efficiency
(C-0.01, F-0.5 ($c=0.1$) and C-0.5).  Note that
F-0.5 is  contained in all three series. These star formation rates were calculated using all
the stars which ended up within $2\ r_{opt}$ of the centre of the final
galaxy.  They thus exclude stars which end up in satellite galaxies.}
\label{sfrs}
\end{figure}

\section{Impact of SN feedback on the formation of galaxies}
\label{results}

In this Section we use our  simulations to investigate
how our model of SN feedback affects the formation of a galaxy
and, in particular, if this process affects the formation
of a disk-like structure.
We also analyse the dependence
of our results on the values of the relevant input parameters.
We will show that SN feedback plays a fundamental
role, regulating the star
formation rate, pressurizing the gas and allowing its late
collapse to a disk structure. 
In the next Section,  we investigate how SN feedback affects
the evolution of two  galaxies with two and three orders of magnitude smaller
mass.

In Table~\ref{simulations_table}
we list a variety of quantities for our simulations:
the masses of dark matter, gas and stars ($M_{\rm DM}$, $M_{\rm gas}$ and $M_{\rm star}$),
the gas fraction  $f_{\rm g}$ (i.e. the  ratio between the gaseous
and baryonic mass), and the baryon fraction $f_{\rm bar}$ 
(i.e. ratio between baryonic and total mass). 
All these quantities are computed at $z=0$ and
refer either to the region within $r_{200}$, the radius
enclosing a mean density $200$ times the critical value, 
or to the  central regions 
($r<2 r_{\rm opt}$, where $r_{\rm opt}$ is defined
as the radius which encloses $83$ percent of the baryonic mass
of the central object). 
Note that our haloes are formed by a central system
(refered to as the central galaxy) and a
series of satellites. Hence,  quantities inside
the virial radius are affected by the properties
of the satellites, whereas those inside twice the optical
radius give a cleaner estimation of the characteristics of the central
system.
In the last column we show 
 $f_*^{z<1}$, which denotes the total
stellar mass formed at $z<1$ divided by the final
stellar mass at $z=0$  (calculated within
$r<2 r_{\rm opt}$).

\subsection{Effects on the star formation process}
\label{sect_sfr}

In Fig.~\ref{sfrs} we show star formation rates (SFR) for our
simulations as a function of time.
We have split our plots  to highlight the dependence
of the results on each of  the input parameters. 
The top panel shows the SFRs for different $\epsilon_c$ values 
(F-0.3, F-0.5 and F-0.9) and for the run without energy feedback (NF).
SN feedback produces a substantial decrease in
the SFR, especially at early times.  In the no-feedback case,
the gas collapses and concentrates at the centre of the potential
well, producing a strong starburst which only fades when the amount
of available gas decreases. Since there is no mechanism to expel
gas in this case, the SFR  behaves
smoothly following the gas collapse.  On the contrary, when SN feedback is considered,
the level of SFR is significantly lower at early times and its bursty behaviour
 indicates  the self-regulation of star formation activity
produced by the injection of SN energy into the interstellar medium.
In this case, each starburst is followed by a depression in the
SFR because the  expansion of surrounding gas 
decreases the amount of material available for star
formation.

Our results are sensitive to the adopted value for the feedback
parameter $\epsilon_c$. 
For larger $\epsilon_c$ values, the cold gas receives more energy, 
and hence we might expect stronger regulation effects.
 Note that  energy
injection to both  hot and  cold phases affects
the evolution of the system. However, the cold gas is
the fuel for star formation, and the effects of
SN feedback on the SFR are more sensitive to energy
deposition in the cold phase.
 Our results indicate that increasing
$\epsilon_c$  reduces the SFR
at early times but can result in more gas being available for
star formation at later phases.

From Table~\ref{simulations_table} we see that
the no-feedback simulation (NF)  forms 
$24.6\times  10^{10}\ h^{-1}$M$_\odot$ of stars
inside the virial radius, and $14.9\times10^{10}\ h^{-1}$M$_\odot$
within $2r_{\rm opt}$. The difference between these two
values reflects  the presence of massive
satellites which  contribute significantly to the stellar 
mass inside $r_{\rm 200}$\footnote{Note that our simulated galaxies are formed by a central system 
and a series of satellites.
In order to have a better
estimation of the contribution of satellites to the stellar
mass of the simulated galaxy, we have done a proper
decomposition  of the mass in the halo into a series of subhaloes,
and calculated the stellar mass contributed by each of these
subhaloes to the total mass. In this way, we find 
that  $25$ percent of the total stellar mass is contributed
by satellites in the final state of the simulation.}. 
As a result of  efficient cooling and star formation,
this simulation ends up with only $22$ percent of the baryons
in the form of gas within the virial radius, and only
$5$ percent in the inner regions.
The inclusion of energy feedback  significantly
changes these numbers, producing a substantial
decrease in the number of stars formed in the galaxy
(see Table~\ref{simulations_table}).
In our feedback cases,
the systems are able to keep
$\sim 50$ percent of the baryons  within the virial radius in the
form of gas at
$z=0$, and from $5$ to $10$ percent in the inner regions.
Note that in the feedback cases, $\sim80$ percent of the stellar mass within
the virial radius is in the main galaxy, so the
satellites  contribute much less to the stellar
mass at $z=0$.

The baryon fraction of the galaxies (listed in 
Table~\ref{simulations_table}) is also affected
by the inclusion of SN energy feedback.
In the no-feedback case (NF),
$f_{\rm bar}^{200}=0.14$ and  $f_{\rm bar}^{2r_{\rm opt}}=0.31$, 
while the feedback cases have
$f_{\rm bar}^{200}=[0.05-0.08]$ and  $f_{\rm bar}^{2r_{\rm opt}}=[0.14-0.17]$ 
which are significantly lower.
This is a consequence of 
galactic winds driven by gas which is heated
by the  SN energy.

The middle panel of Fig.~\ref{sfrs} shows 
SFRs for the experiments with varying energy per SN.
We compare F-0.5 ($E_{\rm SN}=1$), E-0.3, E-0.7 and E-3.
In this case, higher $E_{\rm SN}$ values translate
into lower SFRs and lower final stellar and gas masses
(see Table~\ref{simulations_table}).
In these cases, the gas fraction within the virial radius
increases with the
adopted energy per SN, while the baryon fraction decreases.
These results  are a direct consequence of the higher energy input 
which prevents further star formation and
 produces stronger galactic winds.
The effects of SN feedback are very clear in the extreme
case of very high SN energy ($3\times 10^{51}$ ergs per SN),
where SF occurs in a series of starbursts and its
level is very low. We will show in the next subsection
that this system is not able to form a realistic disk.
 In this case, galactic winds are very
strong, resulting in a low final  stellar mass,
and in a baryon fraction which is 
less than $0.05$, both within the virial radius and
in the innermost region.

Finally, in the lower panel of Fig.~\ref{sfrs} we show SFRs for the
series with different star formation efficiencies ($c$).  In the
absence of SN feedback, a higher $c$ would simply translate into a
higher SFR. However, this is not the case when we consider the
injection of SN energy into the ISM.  Comparing F-0.5 (with $c=0.1$)
with C-0.01 and C0.5, we find that higher $c$ values lead to lower
SFRs and hence lower stellar masses at $z=0$.  This is because
for a given gas distribution, larger $c$ values produce more stars and
so more SN feedback. The surrounding ISM is thus more strongly heated
and reacts by expanding more rapidly. This results in the effective
suppression of further cooling and cuts off the gas supply for further
star formation.  In contrast, a lower $c$ value results in a lower
star formation rate and the reduced feedback heating is unable to
prevent further cooling of gas into the star-forming phase, with the
result that over somewhat longer timescales more stars are formed.  It
is this competition between SN heating and cooling of gas to provide
fuel for star formation which produces the trend between $c$ and the
level of SFR in our simulations. For example, run C-0.01 forms twice
as many stars as run C-0.5.

We have shown that changing the input parameters
of our feedback model produces changes in
the SFRs, the stellar masses,  and the gas and baryon fractions
of the final galaxies. 
Despite these differences, most of our simulations
reproduce approximately the Kennicutt law (Kennicutt 1998) at
$z=0$, with
the exception of our no-feedback model and the most
extreme and least realistic feedback cases   E-3
and C-0.5, which also fail to form disks (see next
subsection).

Self regulation of  star formation 
of the kind found in our models is  crucial  for the
evolution of the galaxies. Gas left over from early times
as a result of  SN energy input can be maintained
in a pressure supported halo which
 later collapses to produce significant star
formation activity in the last 8 Gyr ($z<1$).
In the following  we will see
that this young stellar population is a major contributor to
the formation of disks.

\subsection{Effects on galaxy morphology}
\label{sect_morph}

\begin{figure*}
{\includegraphics[width=55mm]{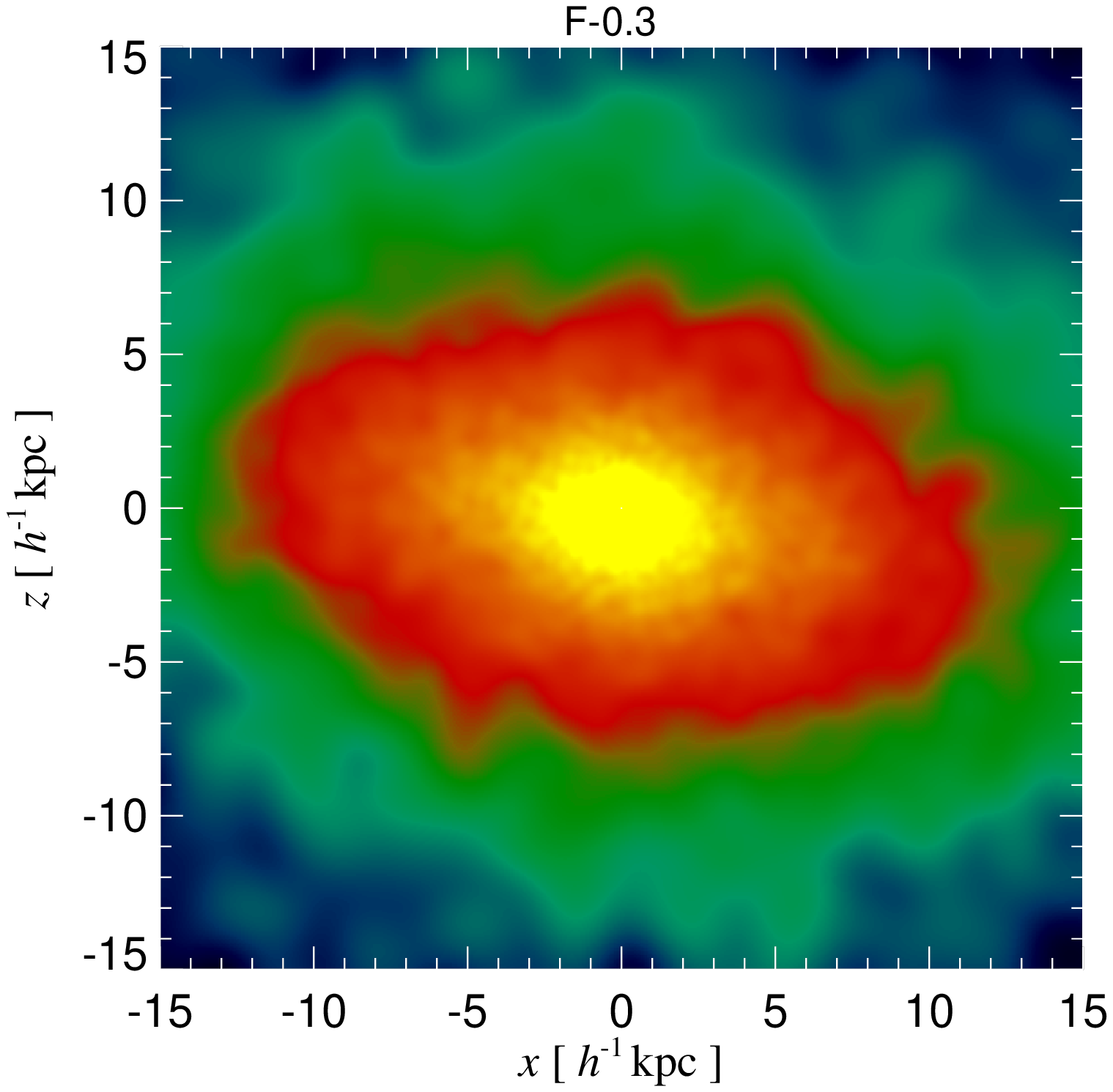}\includegraphics[width=55mm]{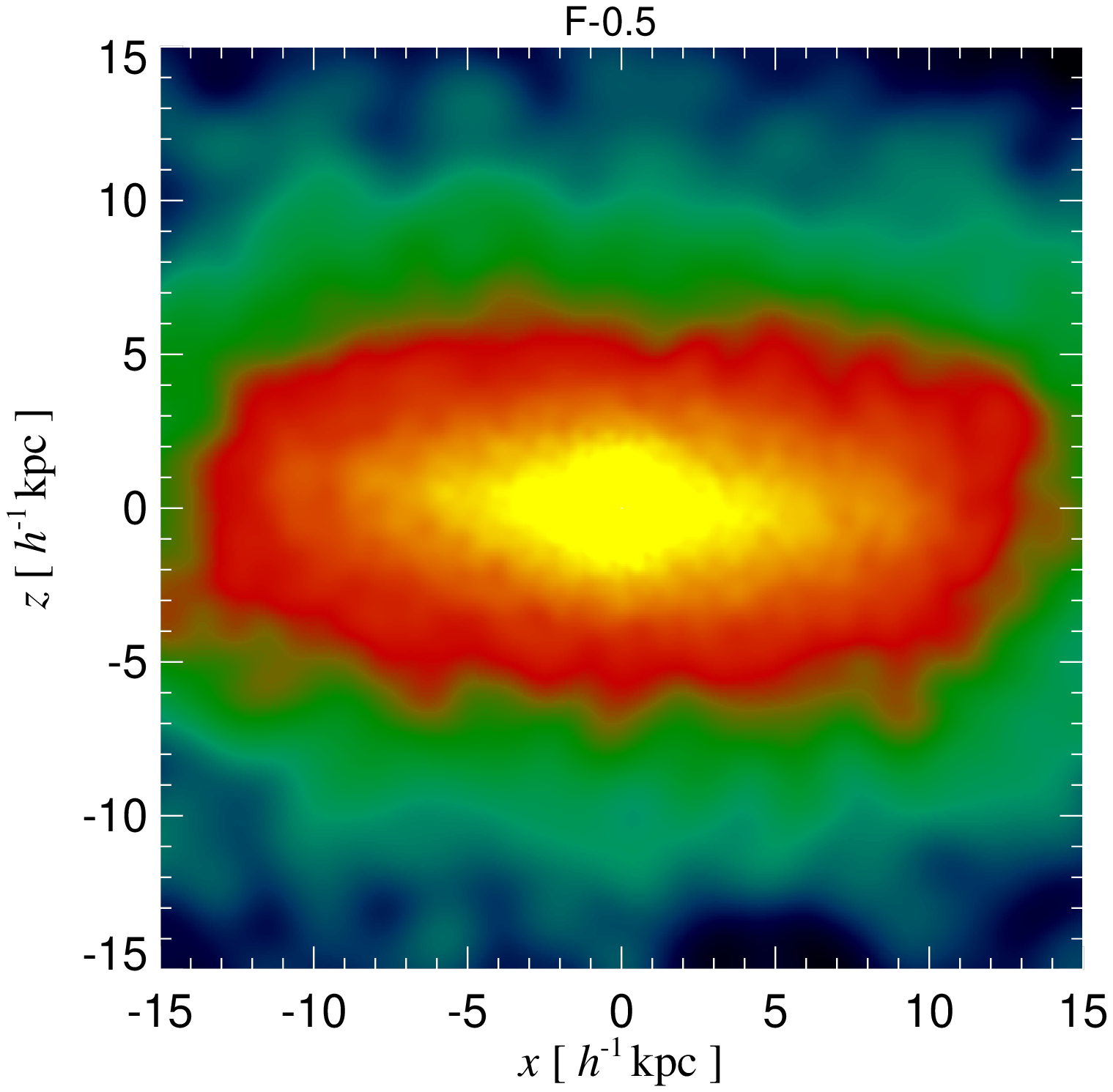}\includegraphics[width=55mm]{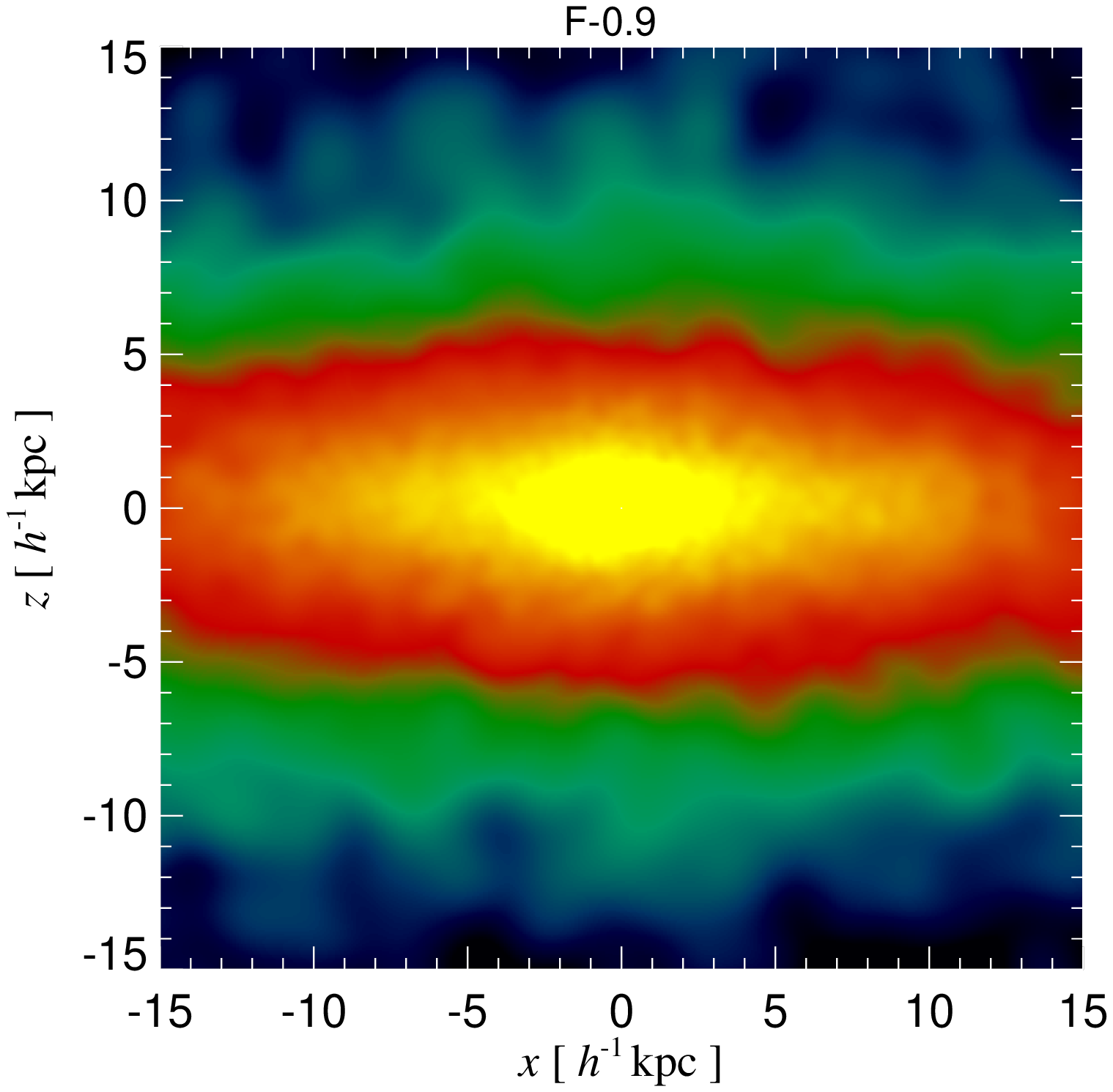}}
{\includegraphics[width=55mm]{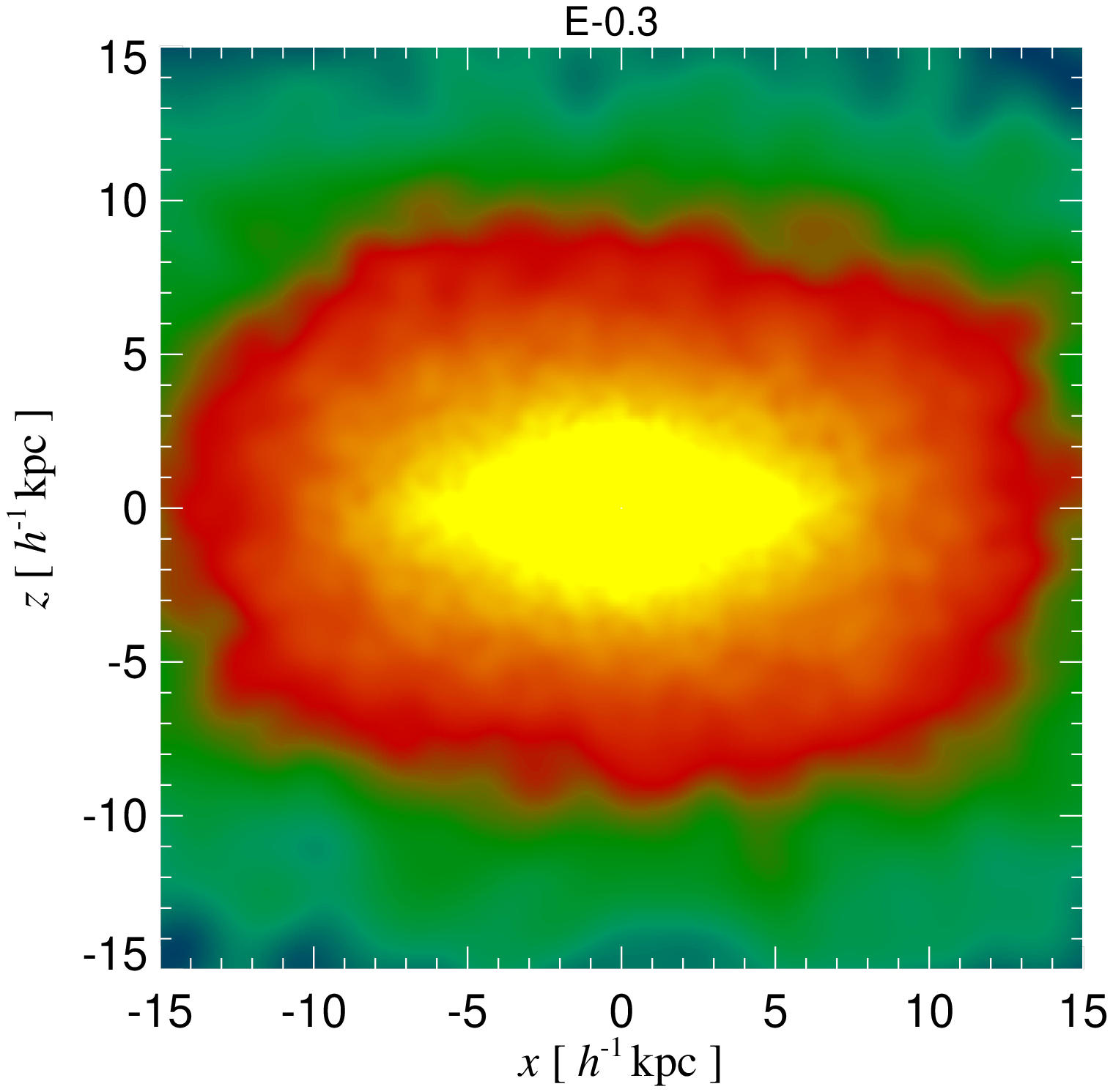}\includegraphics[width=55mm]{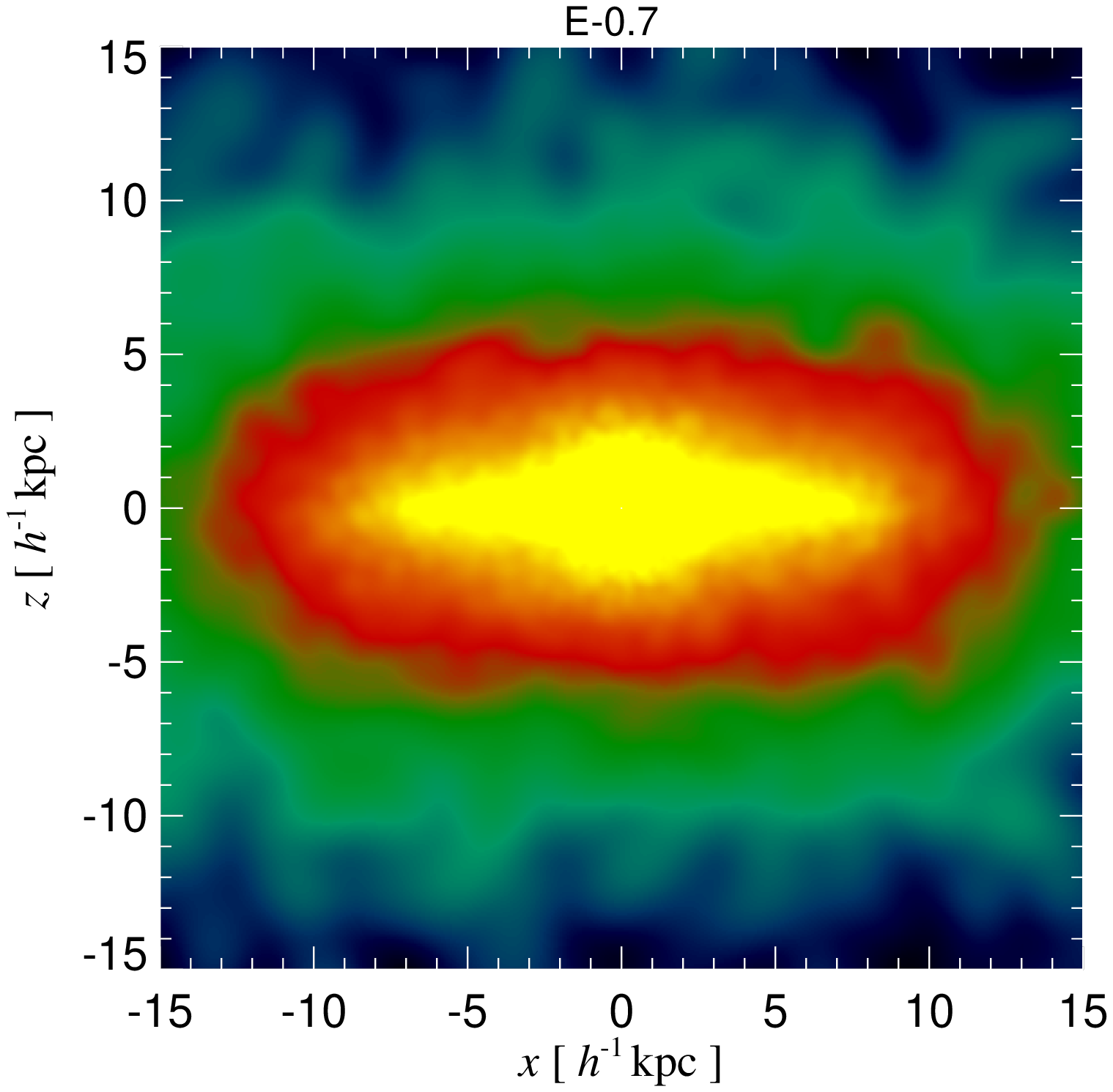}\includegraphics[width=55mm]{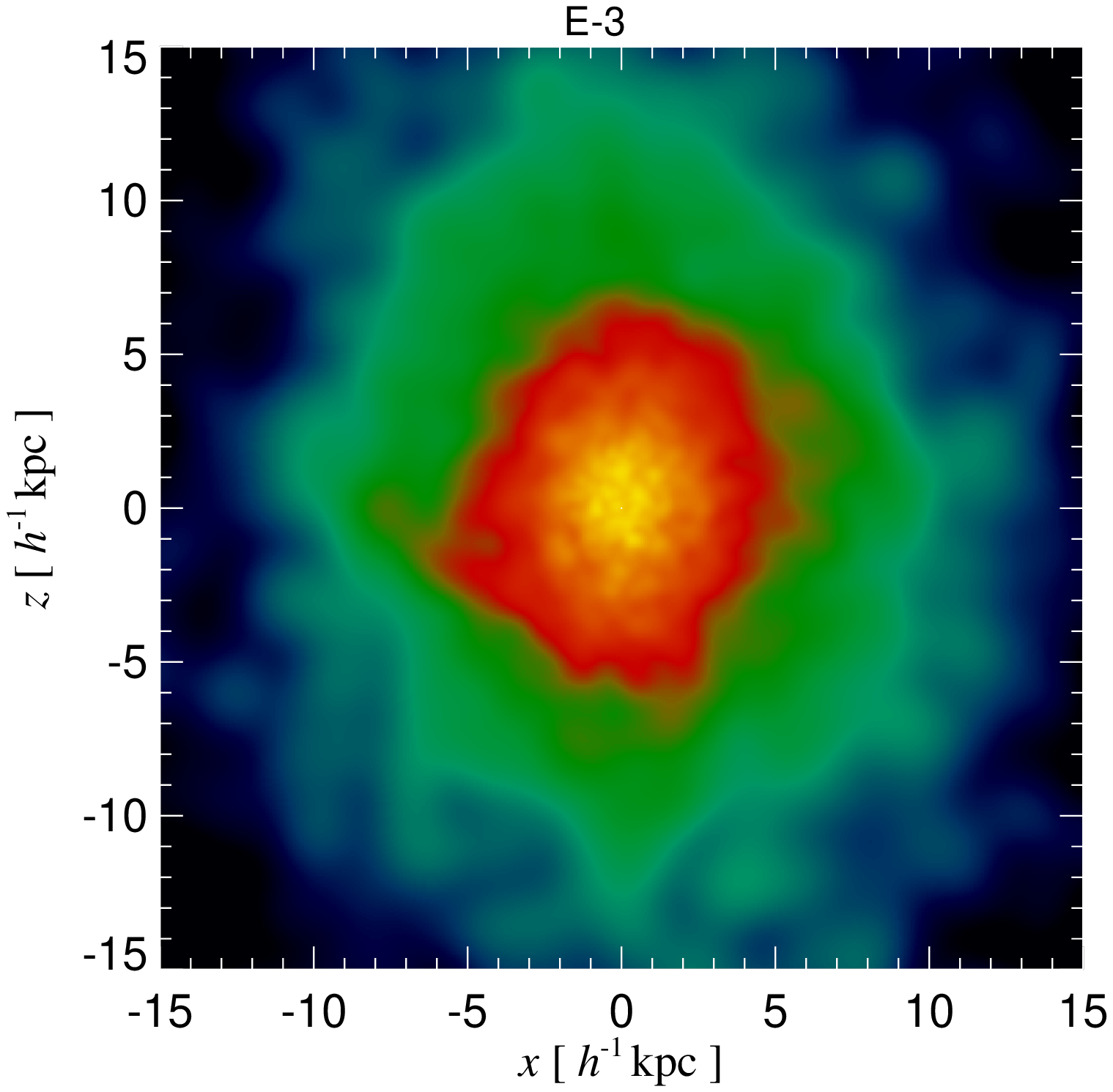}}
{\includegraphics[width=55mm]{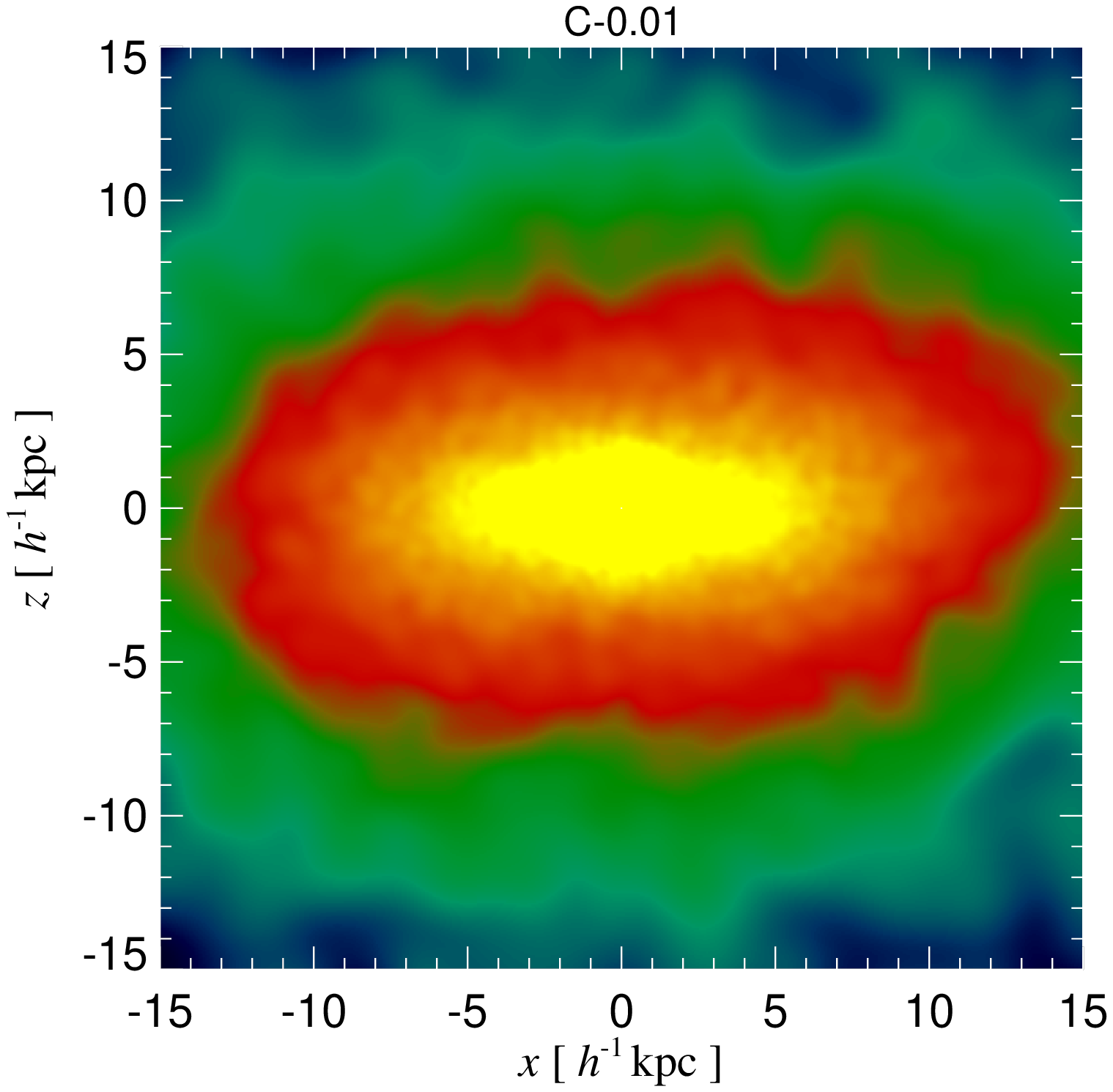}\includegraphics[width=55mm]{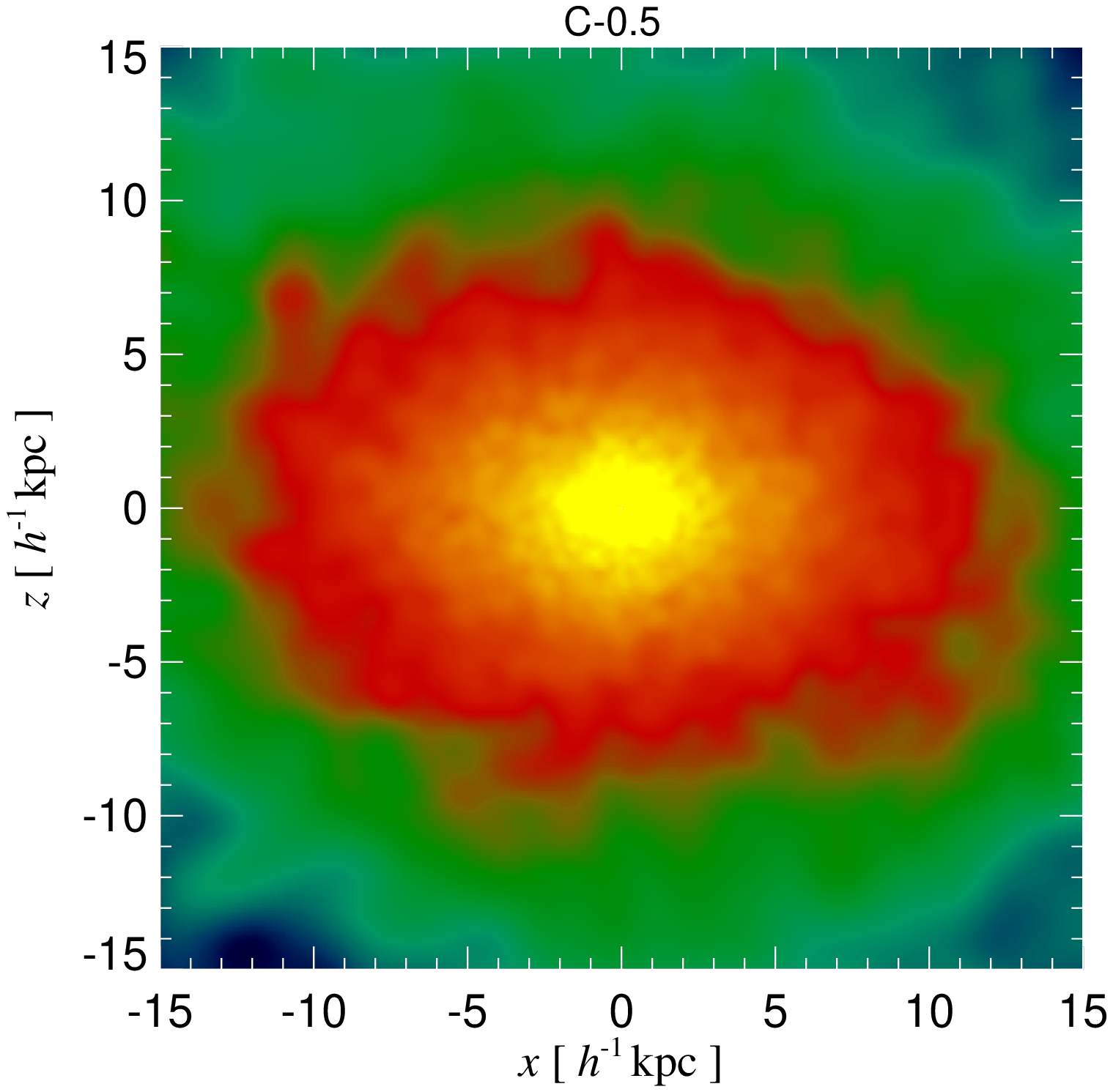}\includegraphics[width=55mm]{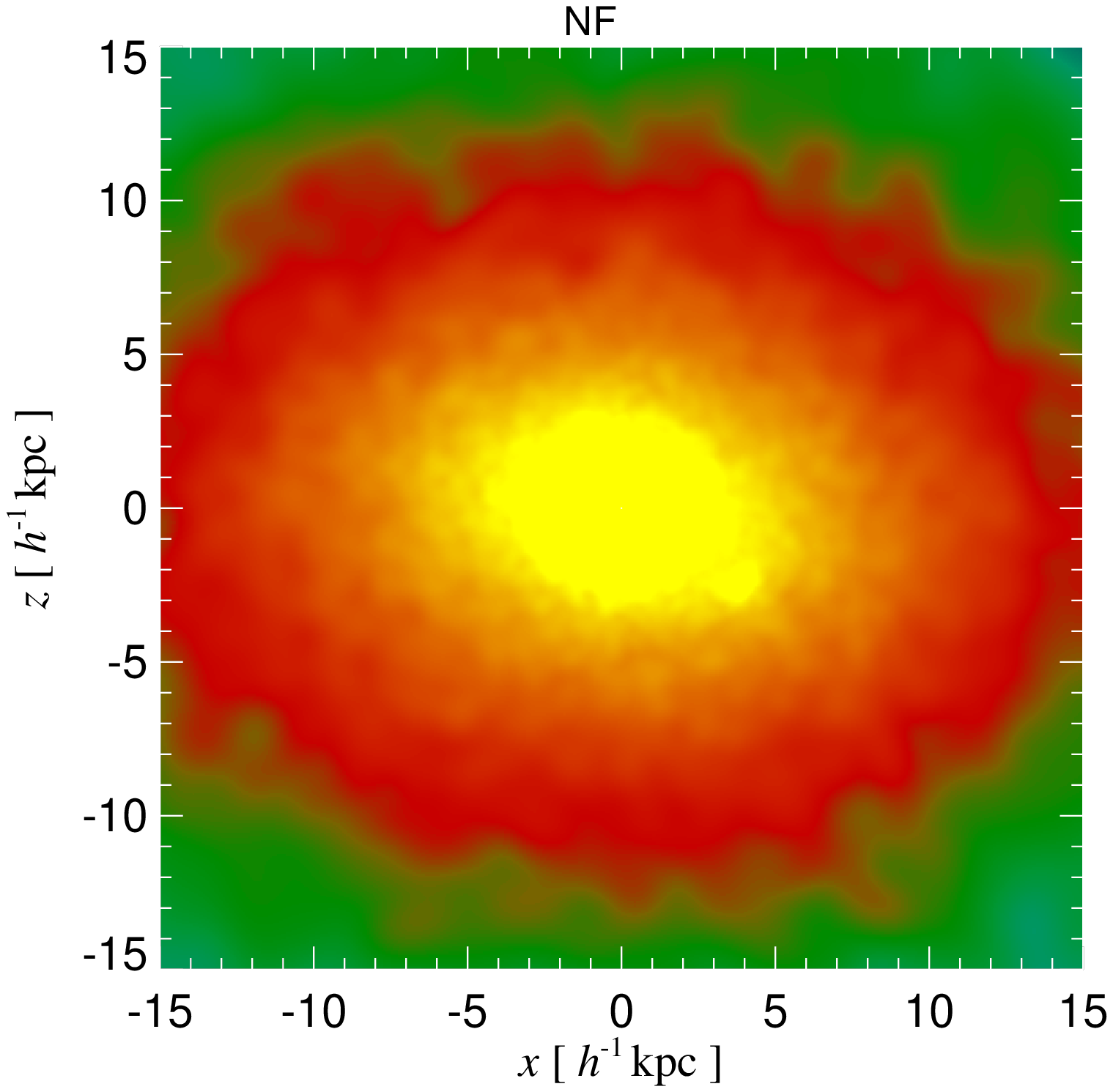}}
\caption{Edge-on stellar surface density maps for $9$ different simulations at $z=0$. The upper panel
shows the results for cases with varying feedback parameter (F-0.3, F-0.5 and F-0.9),
the middle panel the series with different energy per SN (E-0.3, F-0.5 
($E_{\rm SN}=1\times 10^{51}$ ergs) and E-3), and the lower panel the series with
 different star formation efficiencies
(C-0.01, F-0.5 ($c=0.1$) and C-0.5), as well as  the no-feedback simulation (NF).
Note that  F-0.5 belongs to all three series.}
\label{maps}
\end{figure*}

In order to see how SN feedback affects the 
formation of disks, we now examine the
stellar mass distributions of our simulated galaxies.
Fig.~\ref{maps} shows edge-on stellar surface mass density maps 
at $z=0$. 
To make them we 
compute the total angular momentum of the stars in
the central regions ($r<2\ r_{\rm opt}$),
this gives a preferred direction which we define as the $z$-axis
(the angular momentum is  positive with this choice).
The maps were constructed by projecting the mass distributions
within $30$ $h^{-1}$ kpc  in the $xy$ and $xz$ planes. The colors span 4 orders
of magnitude in projected density, with brighter colors
representing higher densities.  
The first row shows the final galaxies for different
feedback parameters ($\epsilon_c$), the middle row corresponds to
different energies per SN ($E_{\rm SN}$) and the lower row shows
galaxies with different star formation efficiencies
($c$),
as well as  the no-feedback case.
Most of our simulations show significant disks,
except for the no-feedback experiment,
E-3  (which is not very realistic
in terms of the energy produced per SN) and C-0.5.

\begin{figure*}
\includegraphics[width=180mm]{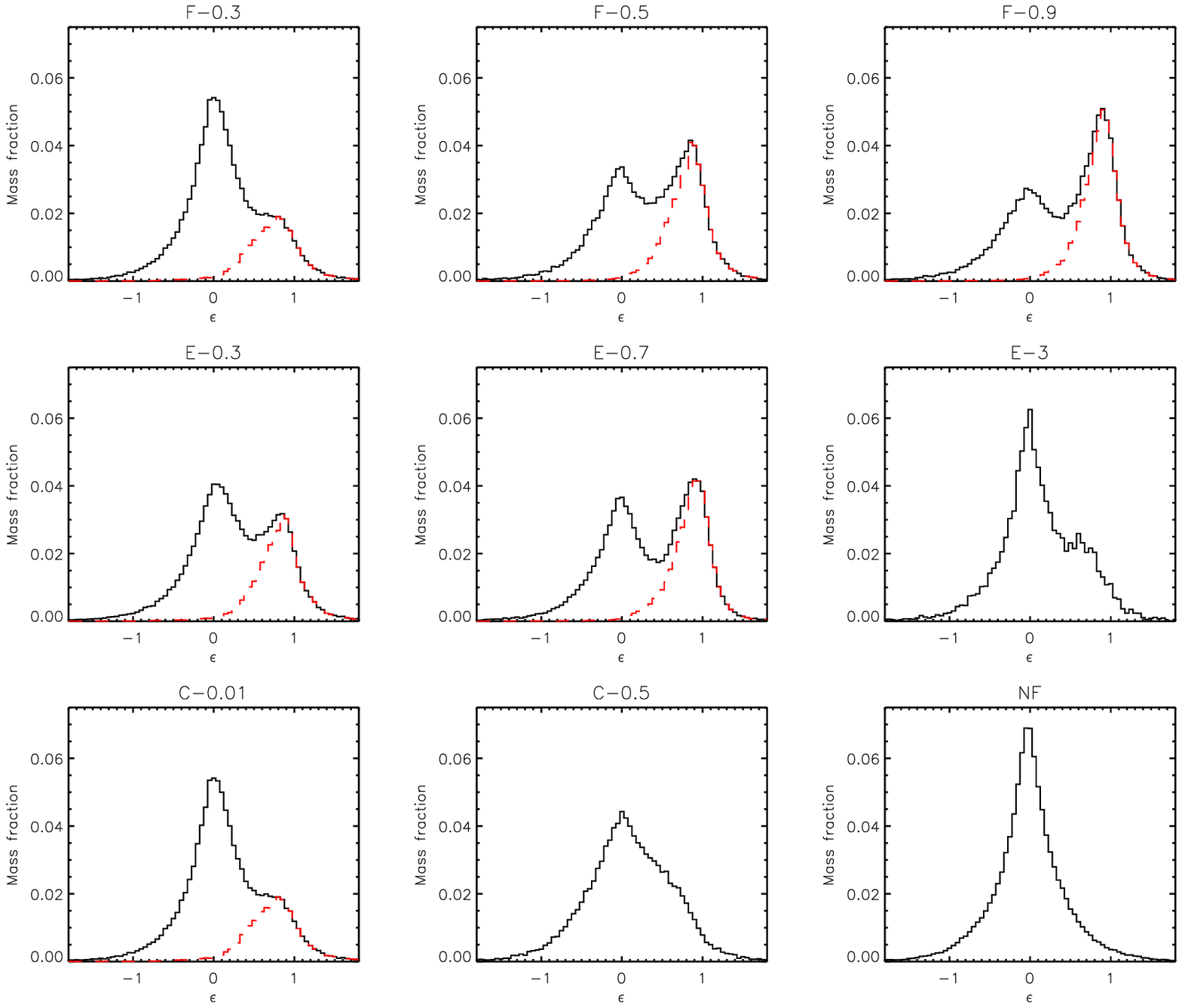}
\caption{Stellar mass fraction as a function of $\epsilon=j_z/j_{\rm circ}$
for the different experiments at $z=0$. The upper panel
shows the results for the series with varying feedback parameter (F-0.3, F-0.5 and F-0.9),
the middle panel the series with different energy per SN (E-0.3, F-0.5 
($E_{\rm SN}=1\times 10^{51}$ ergs) and E-3), and the lower panel the series with
 different star formation efficiencies
(C-0.01, F-0.5 ($c=0.1$) and C-0.5), as well as for the no-feedback
case (NF).
In the cases where a disk-like component is present, we also show
the distribution of $\epsilon$ for disk stars (dashed lines).
In this plot, we only consider stars which belong to the main system
(i.e. discounting stars in satellites).}
\label{hist_epsilon}
\end{figure*}

In order to  quantify the properties of our disks
we have computed the ratio
between the angular momentum of each star in the $z$-direction
 (direction of the angular momentum of stars in
the central region), $j_z$, and
the angular momentum expected for a circular orbit  at the same
radius, $j_{\rm circ}$:
$\epsilon=j_z/j_{\rm circ}$. 
$j_{\rm circ}$ is defined as $j_{\rm circ}= r \cdot v_{\rm
  circ}(r)$\footnote{Note
that our definition of $\epsilon$ differs from that of Abadi et
al. (2004)
and Governato et al. (2007), where $\epsilon$ is defined as 
the ratio between the angular momentum of each star in the direction
of the total angular momentum of the galaxy and the angular momentum
expected for a circular orbit at the same energy.},
where $v_{\rm circ}(r)$ is the circular velocity at radius $r$,
$v_{\rm circ}(r) = \sqrt{G\ M(r) / r}$, given
by the potential well of the system.
A disk supported by rotation is characterized by a population of stars with  $\epsilon\sim 1$.
A component dominated by velocity dispersion
will show a distribution peaked near $\epsilon\sim 0$. 
In Fig.~\ref{hist_epsilon} we show histograms of  $\epsilon$
for our different simulations. 
The distribution in the no-feedback case (NF)
is  consistent with a  pure spheroidal system.
On the contrary, most of our simulations with feedback produce
 significant disks.
Different typical sizes and vertical extensions are found for
the disks when the
input parameters are varied (see Table~\ref{disk_spheroid} where
the characteristic sizes are shown).
Note that these disks coexist with spheroidal components which,
in most cases, dominate the stellar mass of the system.
Although the details of the  mass distributions depend on the input
parameters, it is significant that our model  produces
disks for a wide range of parameters.

\begin{figure*}
{\includegraphics[width=55mm]{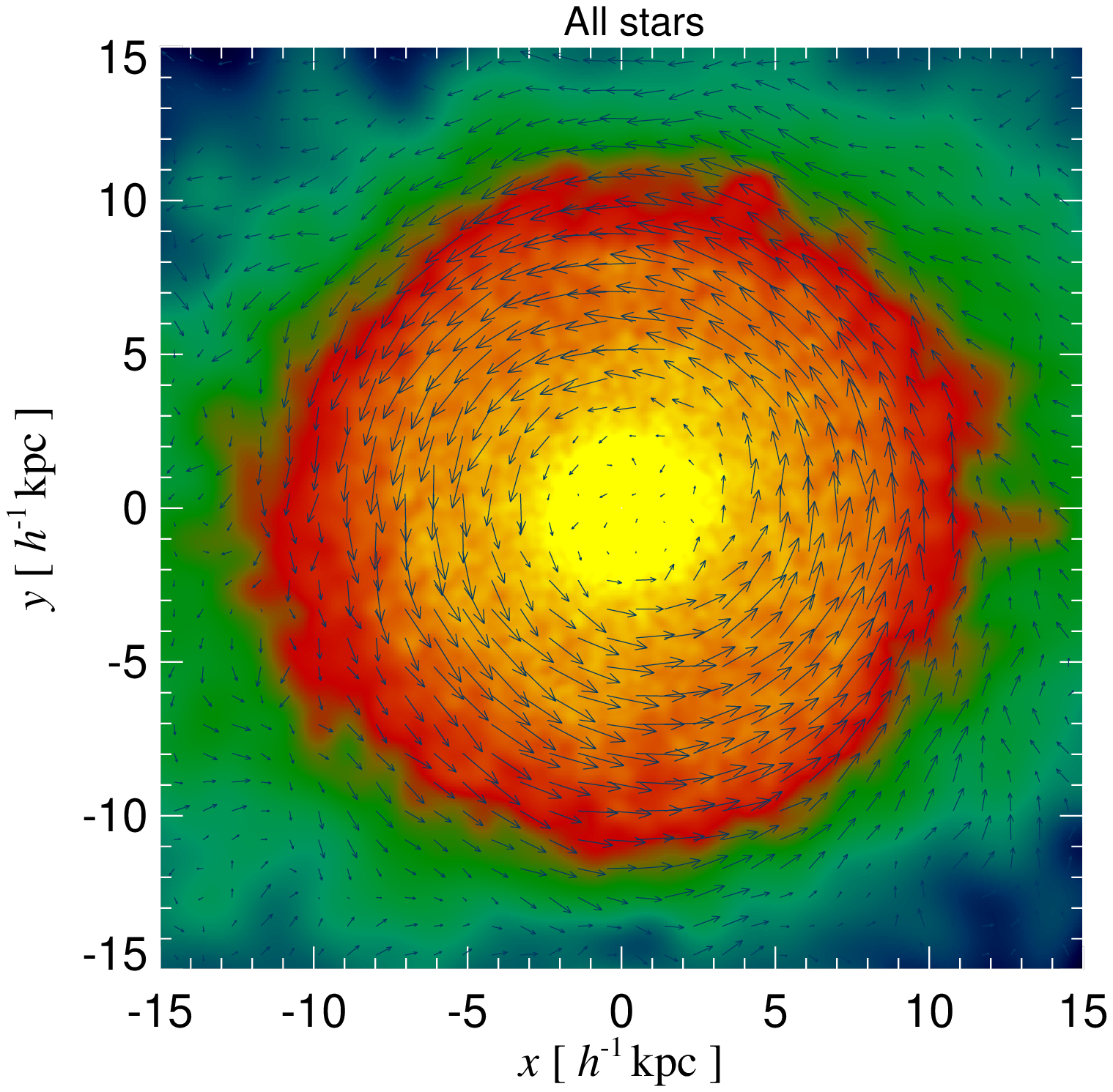}\includegraphics[width=55mm]{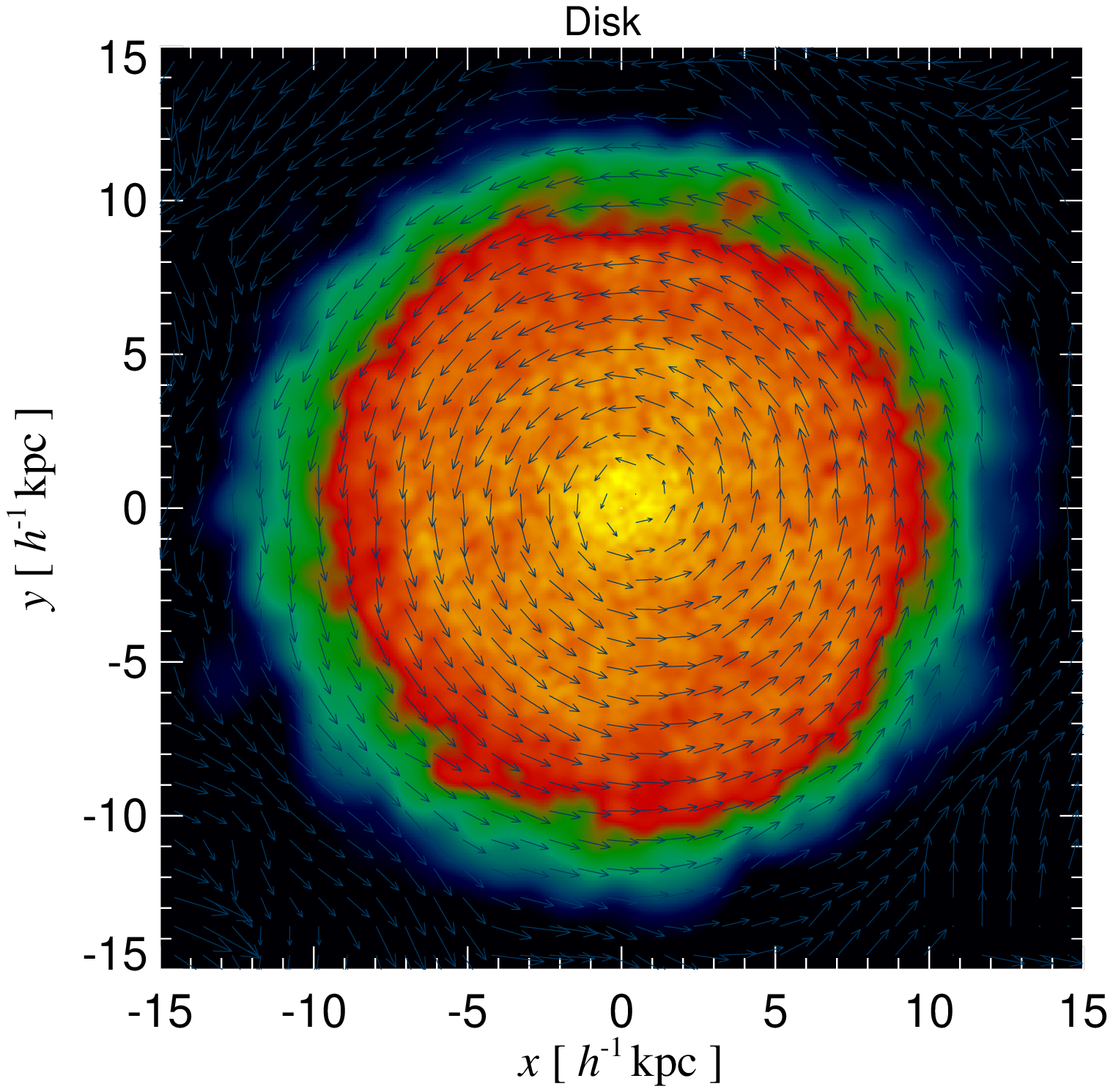}\includegraphics[width=55mm]{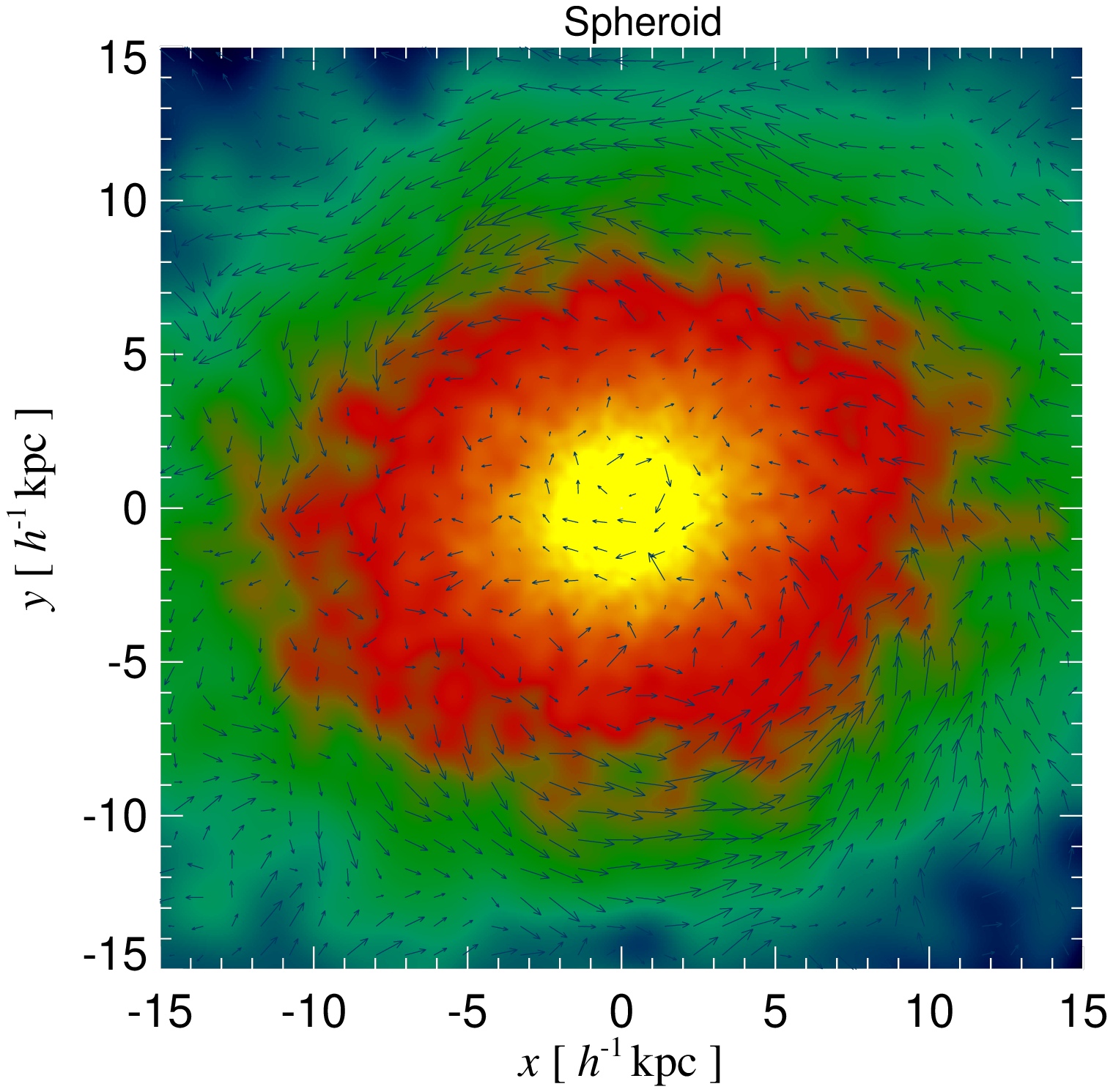}}
{\includegraphics[width=55mm]{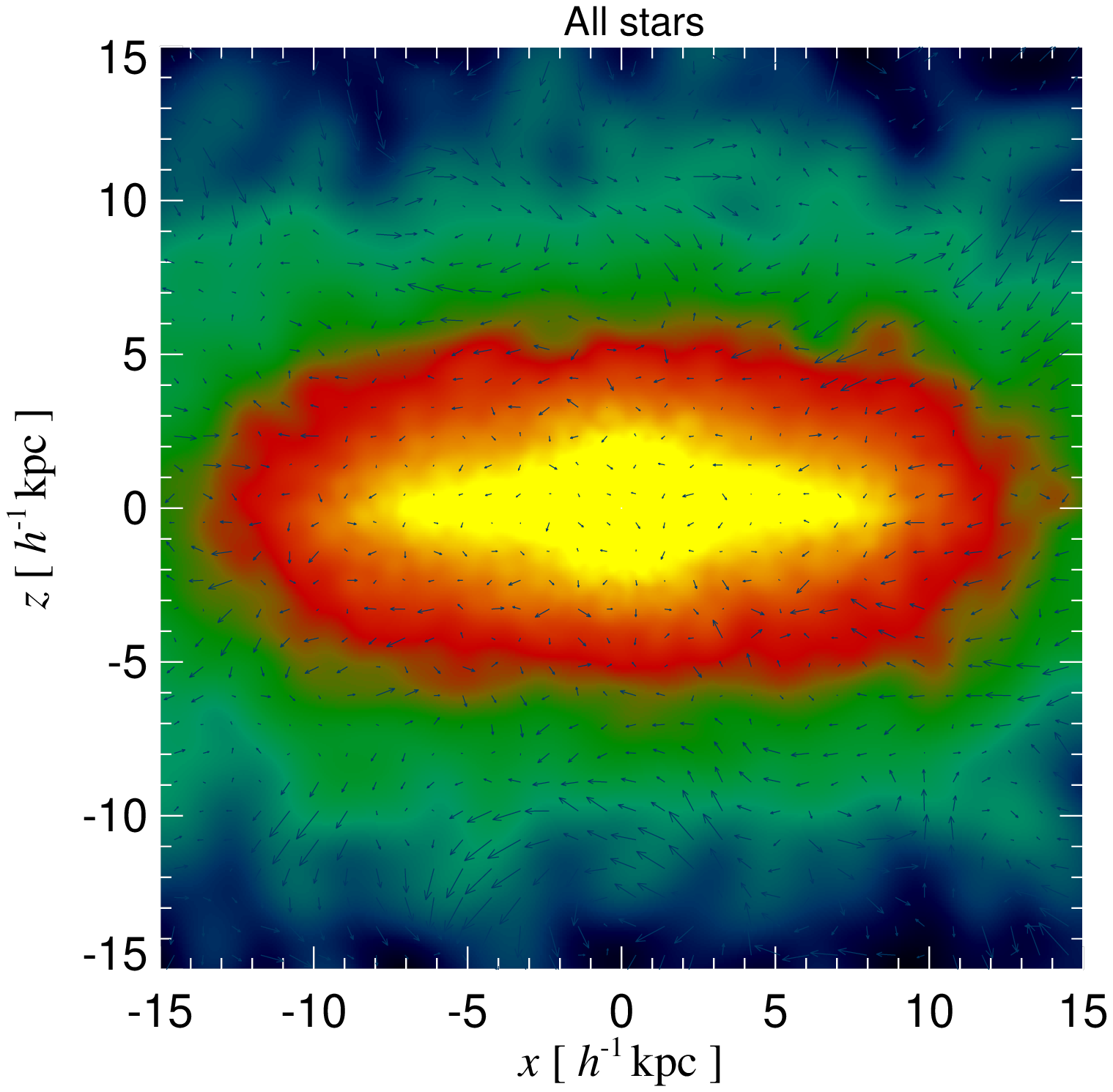}\includegraphics[width=55mm]{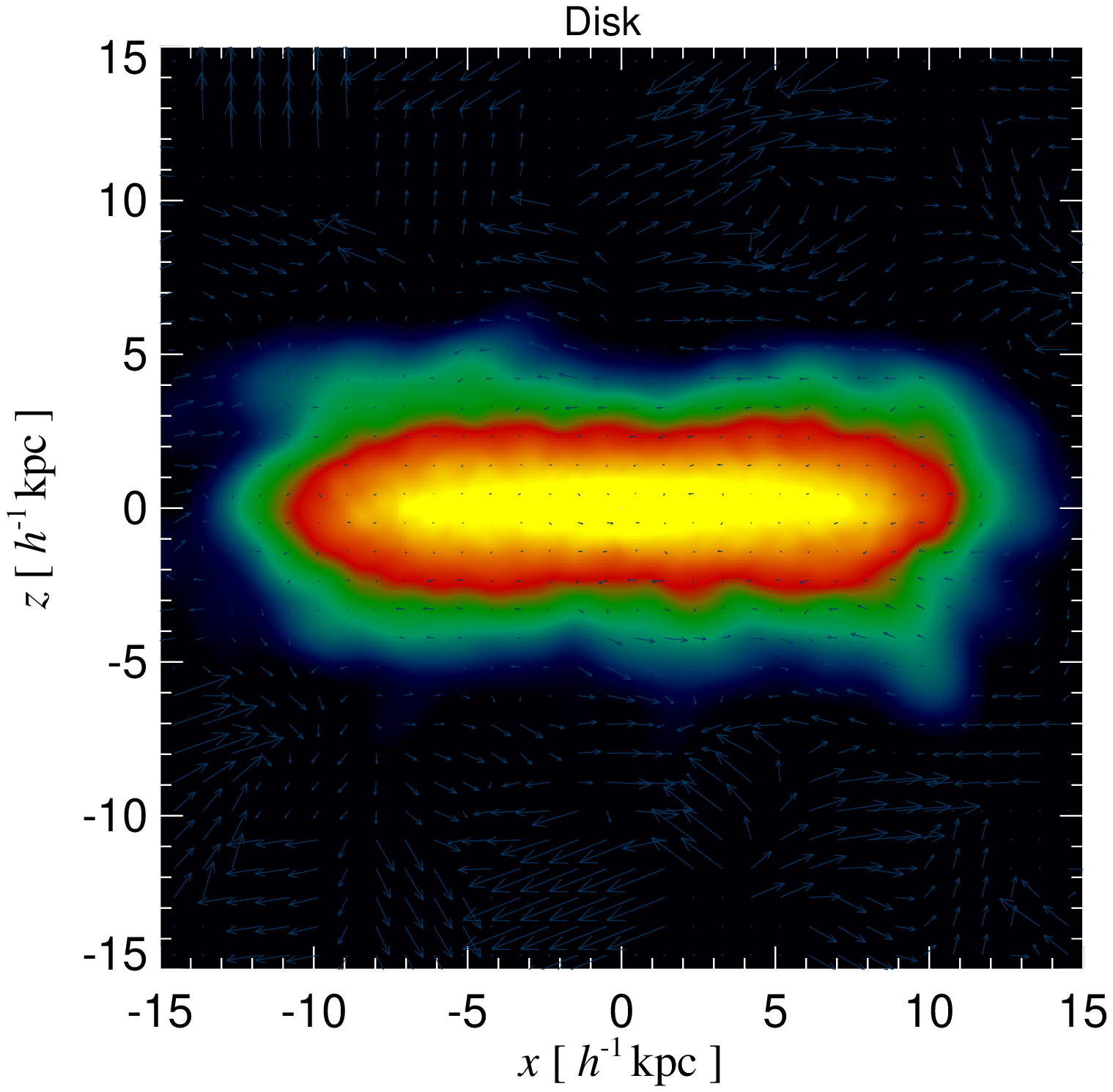}\includegraphics[width=55mm]{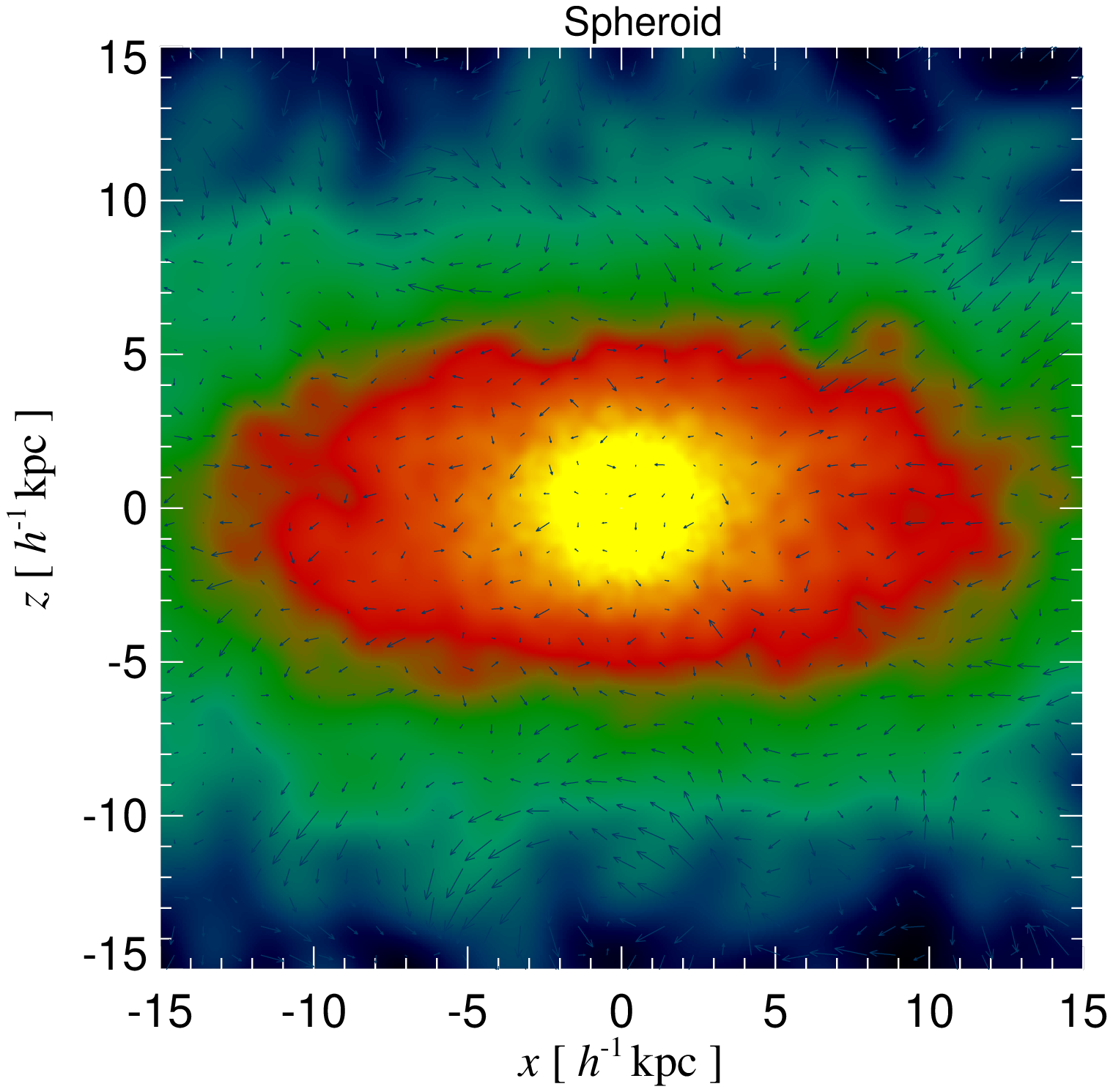}}
\caption{Face-on (upper panels) and edge-on (lower panels) stellar
surface density
maps for the final galaxy formed in E-0.7. The left-hand panel shows the
 complete stellar system, while the middle and right-hand panels
show   views of the disk and spheroidal components,
respectively. These maps all have 
the same color-coding as Fig.~\ref{maps}. The arrows indicate
the mean velocity field of the stars. Their lengths scale with velocity.}
\label{disk_spheroid_plot}
\end{figure*}

We can use our variable $\epsilon$ to quantify the properties of
the disks and to separate the stars into
different components. We have done this  considering
only particles in the main system and neglecting the satellites
to avoid contamination.
We decompose the stars in each simulated  galaxy into two
components which we call {\it disk} and {\it spheroid}.
We refer to the position
of the peak indicative of the disk as $\epsilon_{\rm peak}$.
We assume that stars with $\epsilon\ge\epsilon_{\rm peak}$ are part
of the disk, and generate a symmetric
distribution around $\epsilon_{\rm peak}$. 
For a given star with $\epsilon > \epsilon_{\rm peak}$, we look for a
counterpart with $\epsilon' = 2\epsilon_{\rm peak}-\epsilon$,
and with similar height above the disk plane, similar metallicity and
similar radius. We repeat this for all stars with $\epsilon>\epsilon_{\rm peak}$.
This procedure avoids
ad-hoc parameters and reduces contamination by
spheroid stars.
As an example, Fig.~\ref{disk_spheroid_plot} shows the edge-on and face-on
maps of stellar surface mass  density for the disk and spheroid
components identified in  E-0.7. We also show the corresponding
velocity fields, clearly contrasting the rotation
of the disk with the 
dispersion-dominated spheroid.

We have applied our method to the runs with extended
disk-like components.
In Table~\ref{disk_spheroid} we show the disk and spheroid
masses found for these cases, as well as their ratio.
We list the values corresponding to stars within  $r<2r_{\rm opt}$
of the central galaxy. We also list the half mass radius $r_{\rm disk}$ and half
mass height over the disk plane $h_{\rm disk}$ for the disk component. 
From this table we can see that our experiments have 
a wide range of disk and spheroid masses. In particular,
F-0.5, F-0.9 and E-0.7 show the largest ratios between
disk and spheroid mass. In the case of F-0.9, 
the disk and spheroidal components have the same mass,
in all other cases the spheroid dominates in terms of
mass (but may not dominate if we consider luminosities\footnote{
By combining the information on the age and metallicity of the
stars, we have calculated the disk and bulge ($r<5\ h^{-1}$ kpc) luminosities
using the GALAXEV code (Bruzual \& Charlot 2003).
This yields a disk-to-bulge ratio of 2.4 in the R-band (and
a bulge-to-total ratio of 0.30).
This is
low with respect to observational estimates for
Milky-Way type galaxies, particularly in view of the fact that we
neglect the effects of dust (e.g. Gadotti 2008).}).
Thus, changing the input parameters
leads to a variety of  different disk sizes and 
 scale lengths.

In Table~\ref{disk_spheroid}
we show the time when half of the mass of each
component was assembled ($\tau_{\rm disk}$ and $\tau_{\rm spheroid}$).
In the no-feedback case,  the typical
scale of star formation is very short, with $\tau_{\rm spheroid}=1.17$ Gyr,
and the spheroidal components in the feedback cases
are also mostly old, with characteristic formation times below $3$ Gyr.  
In the case of the disk components in the feedback experiments,
their half mass times indicate that they are
young, with typical $\tau_{\rm disk}$ of $6$ Gyr.

Our star formation and feedback model clearly can produce disks
in a cosmological context for any choice
of input parameters 
but spheroid mass dominates over disk mass in all cases
in this particular halo, our best case being a 1:1 disk-to-spheroid mass.
In the following subsections we choose one of our
simulated galaxies and compare its evolution with
that of the galaxy formed in the no-feedback case, in
order to investigate in more detail how disks form and
how SN feedback affects the  distributions of specific
angular
momentum, stellar age and chemical abundance.

\begin{table*}
\begin{small}
\caption{Mass of disk and spheroidal components 
(in units of  $10^{10}\ h^{-1} \ M_\odot$) and ratio between disk and
spheroid masses ($D/S$)  for the different  simulations. We also show the disk half
mass radius ($r_{\rm disk}$) and half mass vertical scalelength
($h_{\rm disk}$) in  $h^{-1}$ kpc, as well as the
time when half of the final mass of the disk was formed
($\tau_{\rm disk}$) in Gyr, as well as the corresponding time for 
the spheroid ($\tau_{\rm spheroid}$).
We also show the fraction of metals in the stellar ($f^{\rm met}_{\rm
  stars}$) and gas
($f^{\rm met}_{\rm gas}$) components
and, for the gas component, the fraction of metals within $2\ r_{\rm
  opt}$ ($f^{\rm met}_{\rm gas, in}$) and between  $2\ r_{\rm
  opt}$ and the virial radius ($f^{\rm met}_{\rm gas, out}$).}
\vspace{0.1cm}
\label{disk_spheroid}
\begin{center}
\begin{tabular}{lccccccccccc}
\hline
Test  &   $M_{\rm disk}^{2\rm{r_{\rm opt}}}$ & $M_{\rm spheroid}^{2\rm{r_{\rm opt}}}$ &
 $D/S$ & $r_{\rm disk}$  & $h_{\rm disk}$  & $\tau_{\rm disk}$ &
 $\tau_{\rm spheroid}$ & $f^{\rm met}_{\rm stars}$ & $f^{\rm met}_{\rm gas}$
& $f^{\rm met}_{\rm gas, in}$ &$f^{\rm met}_{\rm gas, out}$ \\\hline
NF     &   -   &  14.4  & -     & -    & -    & -    & 1.17 & 0.99 &
0.01 &  -  & - \\
F-0.3  &  1.29 &  3.07  & 0.42  & 5.82 & 1.14 & 4.75 & 2.95 &0.31 &
0.69 & 0.26 & 0.74 \\ 
F-0.5  &  2.46 &  3.00  & 0.82  & 6.44 & 0.84 & 5.54 & 2.78 & 0.22 &
0.78 & 0.29 & 0.80 \\ 
F-0.9  &  3.35 &  3.23  & 1.04  & 9.74 & 0.98 & 6.24 & 2.33 & 0.32 &
0.68 & 0.22 & 0.78 \\ 
E-0.3  &  4.85 &  8.01  & 0.60  & 4.75 & 0.66 & 5.72 & 2.54 & 0.28 &
0.72 & 0.28 & 0.72 \\
E-0.7  &  3.33 &  4.05  & 0.82  & 5.72 & 0.50 & 6.33 & 2.49 & 0.20 &
0.80 & 0.37 & 0.63 \\
E-3    &   -   &  1.23  &  -    & -    & -    &   -  & 2.64 & 0.15 &
0.85 & 0.13 & 0.87\\ 
C-0.01 &  2.71 &  7.05  & 0.39  & 2.64 & 0.40 & 4.78 & 2.83 & 0.17 &
0.83 & 0.21 & 0.79 \\ 
C-0.5  &   -   &  4.36  &  -    & -    & -    &   -  & 2.38 & 0.25 &
0.75 & 0.16 & 0.84\\\hline

\end{tabular}
\end{center}
\end{small}
\end{table*}

\subsection{Evolution of the angular momentum}
\label{sect_j_age_fe}

\begin{figure*}
\includegraphics[width=90mm]{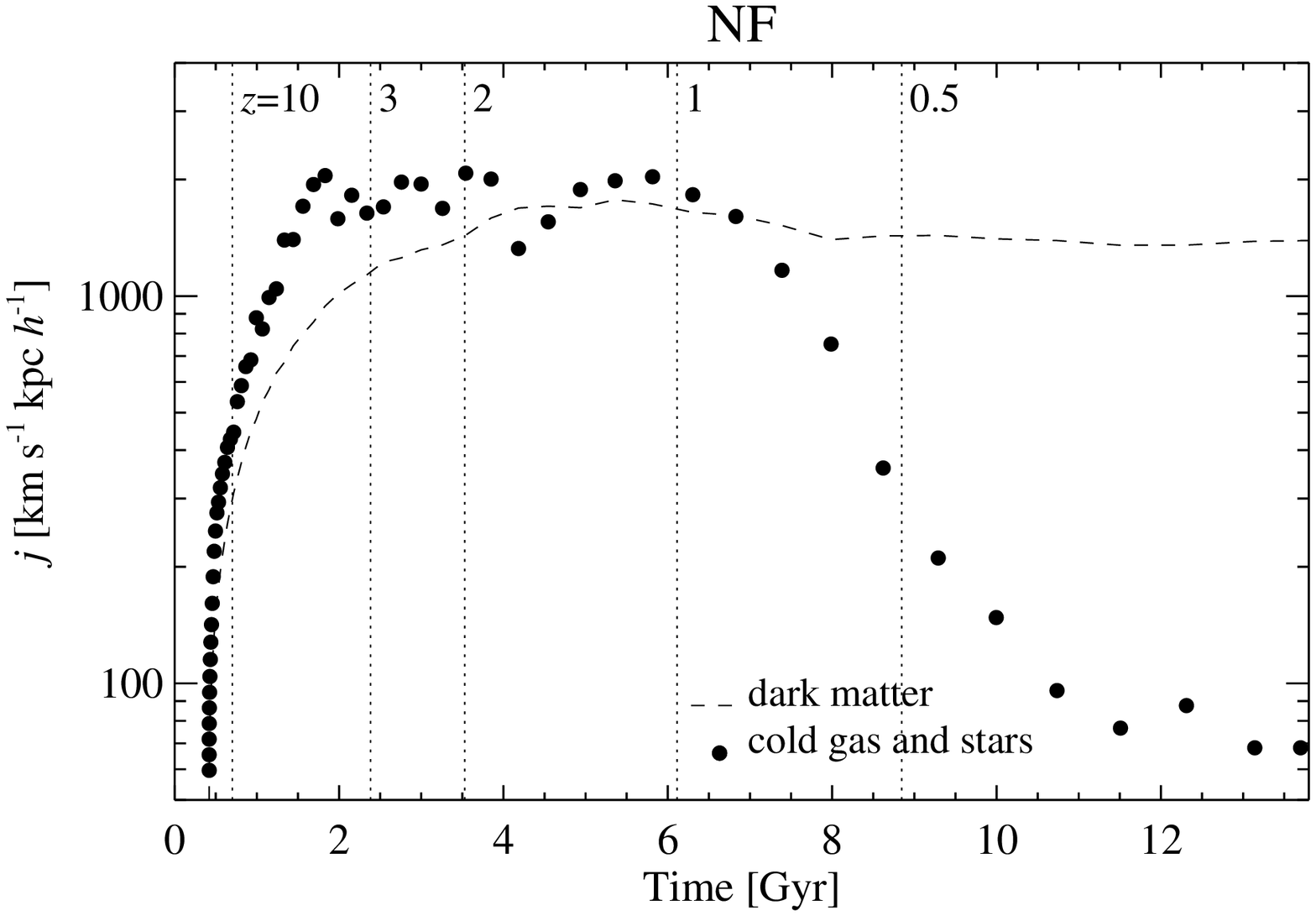}\includegraphics[width=90mm]{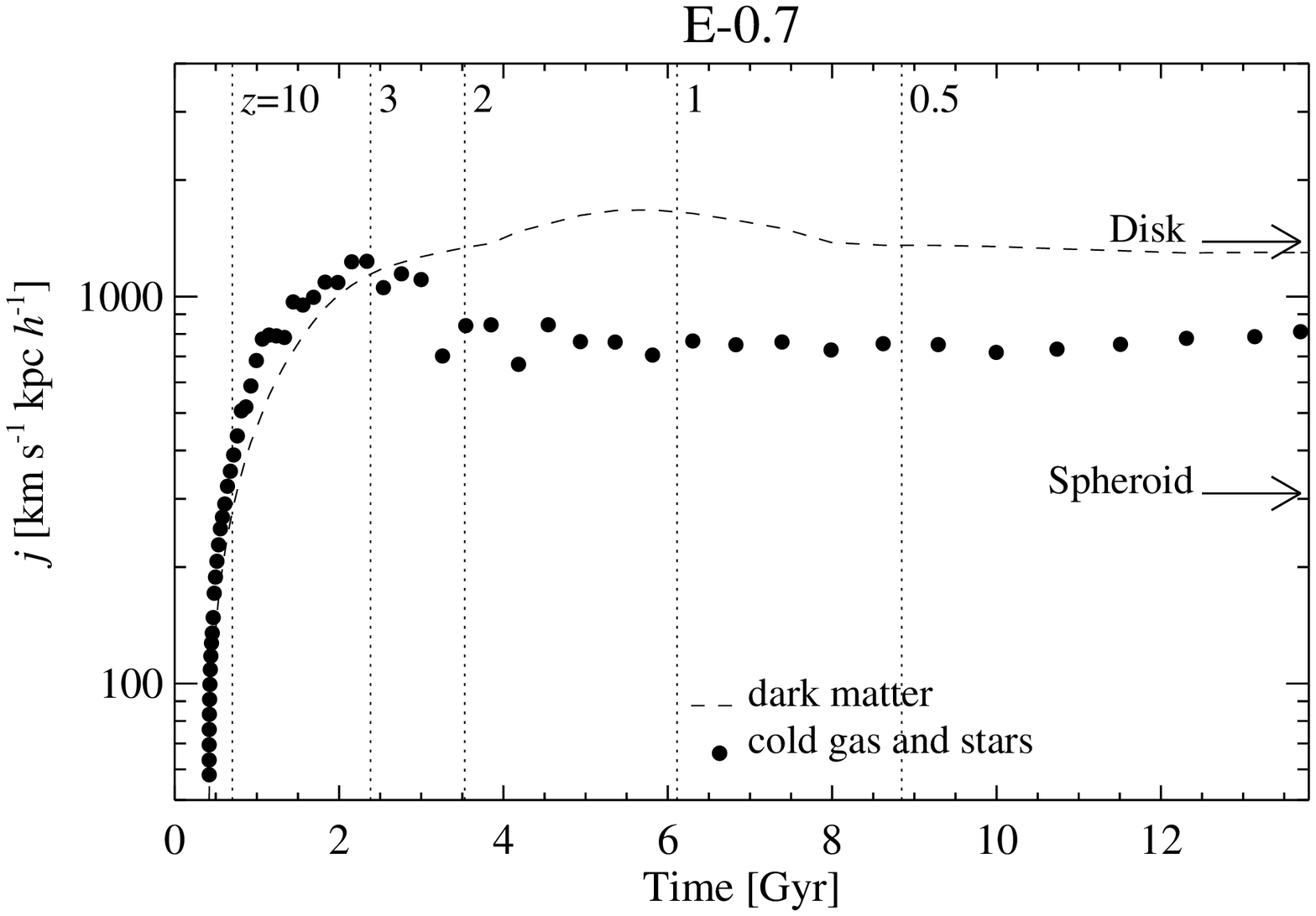}
\caption{Dashed lines show the specific angular momentum as a function of time 
for the dark matter that, at $z=0$, lies within the virial radius of
the system for  NF (left panel) and E-0.7 (right panel).
We also show with dots the specific angular momentum for the baryons
which end up
as cold gas or stars in the central $20\ h^{-1}$ kpc at $z=0$.
The arrows show the specific angular momentum of
disk and spheroid stars. }
\label{specific_J}
\end{figure*}

Our model E-0.7 has  one of the largest ratios of disk to
spheroid mass, and has the thinnest disk. 
We have selected this simulation as an example to study
in more detail the effects of SN feedback in galaxy evolution.
The results for this particular simulation are qualitatively similar
to those for all other cases where disk-like components  formed.
In  order to investigate why a disk forms with feedback but not in 
the no-feedback case (NF) we compare the
evolution of the specific angular momentum
(i.e. angular momentum per unit mass)  in the two
simulations. 
We have computed the specific angular momentum
for the dark matter component in the no-feedback and feedback  simulations, using particles
which ended up within the virial radius at $z=0$. We followed these
particles back in time, calculating their centre of mass and
the corresponding specific angular momentum. 
We have excluded $z=0$ satellites which contribute
an important fraction of the total specific angular momentum
at the end,
especially in the no-feedback case.
We have also calculated the corresponding specific
angular momentum for the cold gas plus stars which ended up 
within $2\ r_{\rm opt}$ at $z=0$. 
In this way, we can
investigate the relation between the formation of the
disk and the evolution of the specific angular momentum of the
galaxy.

In  Fig.~\ref{specific_J}
we show  results for the evolution of the specific angular momentum
of the dark matter and of the cold gas plus stars, calculated as
explained
above, the left-hand panel shows results for the no-feedback case whereas the
right-hand
panel corresponds to E-0.7. 
These plots show that the evolution of the 
specific angular momentum of the dark matter  component
is similar in the two cases, growing as a result of tidal torques
at early epochs and being conserved from turnaround ($z\approx 1.5$)
until $z=0$.

The specific angular momentum of the cold gas plus stars, however,
behaves differently in the no-feedback and feedback experiments, in
particular at late times.
During the early stages of  evolution, 
the  specific angular momentum of this component also grows through
the
tidal torques exerted by nearby protogalaxies
and it maximises at a similar value in the two simulations which is
close to
the maximum value for the dark matter.
However, at later times, the specific angular momentum of the baryonic
component
behaves differently in the two cases.
In the no-feedback case (NF), 
much angular  momentum is lost through dynamical friction, particularly
through a  satellite which is accreted onto the main halo at $z\sim 1$. 
The decrease in specific angular momentum
after this time is more than an order of magnitude.
In E-0.7, on the other hand,  the cold gas
and stars lose rather little specific angular momentum between $z=1$ and $z=0$.
Two main factors contribute to this difference.
Firstly, in E-0.7  a significant number of
young stars form between $z=1$ and $z=0$
with high specific angular momentum (these stars form from
high specific angular momentum gas which becomes cold at late times); and secondly, 
dynamical friction affects the system much less than in NF,
since satellites are less massive.
Our results are similar to those  of
by Zavala, Okamoto \& Frenk (2007), who
analysed two simulations with identical initial
conditions but differing baryonic physics, one
of them resulting in a disk-dominated galaxy,
the second  in a spheroidal system.

Finally, in the right panel of Fig.~\ref{specific_J} we
show  the final specific angular momentum
of the disk and spheroidal components identified in this
simulation at $z=0$.
Clearly the disk component has a high
specific angular momentum, comparable to that of
the dark matter, while the spheroid formed in
this case has a much lower specific angular momentum.
This reflects the close relation between angular momentum
and morphology.

\subsection{Stellar age distributions}
\label{stell_age}

Our results suggest that the formation of the
disks is closely related to the ability to form
stars at late times. In our no-feedback
experiment, star formation occurs early,
giving rise to a stellar spheroid at $z=0$ which is dominated
by old stars.
On the contrary, in E-0.7 the
stellar component has a larger contribution
from young stars. This behaviour is found in
 all simulations with feedback, as can be seen from 
Table~\ref{simulations_table} where we show
the fraction of the final stellar mass formed
between $z=1$ and $z=0$, $f_*^{z<1}$. The no-feedback
simulation has the lowest value for $f_*^{z<1}$,
while the largest values are found for the
systems with the most massive disks.

\begin{figure}
\includegraphics[width=80mm]{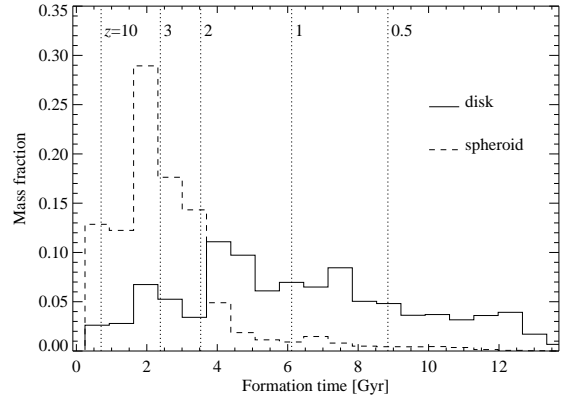}
\caption{Mass fraction as a function of formation time for stars
of the disk and spheroidal components in simulation E-0.7.}
\label{hist_stellarage}
\end{figure}

In order to analyse in more detail the relation between
disk formation and delayed star formation,
we show in Fig.~\ref{hist_stellarage}  the distribution of
stellar formation times for the disk and spheroidal components in  E-0.7. 
The two components
differ significantly in their  formation times, the spheroid
being composed of old stars while the disk contains a younger stellar population.
In particular, only $6$
percent of the stars in the spheroidal component
were formed at times later than $6$ Gyr ($z\approx 1$).
On the contrary, 
$53$ percent of the final stellar mass
of the disk was formed between $z=1$ and $z=0$,
$24$ percent in  $z=[0.5,0]$, and $29$
percent in $z=[1,0.5]$.
This suggests that the ability of a system
to form a disk  correlates with its ability to
retain enough gas to form stars
at recent times where mergers are not as frequent as
at early epochs. This
 becomes possible when SNe regulate
 star formation activity.
We note that in our simulation without SN
energy feedback (NF), only $10$ percent of the final stellar
mass forms after $z=1$ (see Table~\ref{simulations_table}).

The age of the disk components can also be quantified
by the time when half of the final disk stars had
already formed,  $\tau_{\rm disk}$. The $\tau_{\rm disk}$ values  
for the different simulations
are shown in Table~\ref{disk_spheroid}. It can be seen
 that large disks typically have
 larger  $\tau_{\rm disk}$ values.
In Table~\ref{disk_spheroid} we also show the
half mass formation time for the spheroidal components,
 $\tau_{\rm spheroid}$. We find that, in all cases, 
the spheroids are  mainly old. Note that the no-feedback case has the
oldest stellar population, while all the feedback
cases, independent of the disk they were able
to form, show similar values of $\tau_{\rm spheroid}$.

\begin{figure}
\includegraphics[width=80mm]{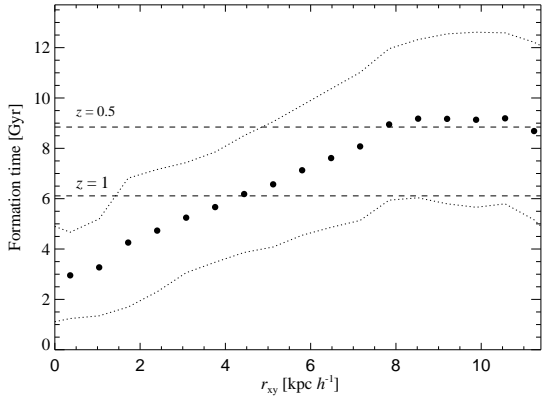}
\caption{Formation time of disk stars within $2\ r_{\rm disk}$ as a function of distance 
within the disk plane
for the feedback simulation E-0.7. The dotted lines give the
$\pm\sigma$ scatter  and dashed lines depict the times
corresponding to $z=1$ and $z=0.5$.}
\label{inside_out}
\end{figure}

In order to investigate the spatial history of disk assembly,
 we plot in Fig.~\ref{inside_out} the mean formation time
of disk stars  as a function
of radius within the disk plane for E-0.7. We also show
the $1\sigma$ scatter  and we note the times
corresponding to $z=1$ and $z=0.5$. As noted above, $29$ and $24$ percent of the final 
stellar mass of the disk was formed in the redshift ranges
$z=[1,0.5]$ and  $z=[0.5,0]$, respectively.
Clearly the stellar disk 
formed from the inside-out, although the dispersion
is high.
In fact, we find that star formation between $z=1$ and $z=0$
lies primarily in a ring-like structure whose radius increases
with time.
In contrast, the spheroidal component shows no trend
between the age of stars and radius.

\subsection{Chemical abundances}
\label{chem_prop}

In this section we investigate the  chemical
properties of our simulated galaxies. 
Metals are produced in stellar interiors and ejected
into the interstellar medium when supernova explosions take
place. If feedback is neglected,
most metals remain locked 
either in the stars or in the cold gas near
star forming sites, since there is no efficient
mechanism to transport gas and 
metals from the inner to the outer regions.
Winds generated when SN feedback is included
produce a redistribution of mass and metals,
and the resulting chemical properties are very
different from those obtained when SN feedback is not
considered.

\begin{figure*}
\includegraphics[width=80mm]{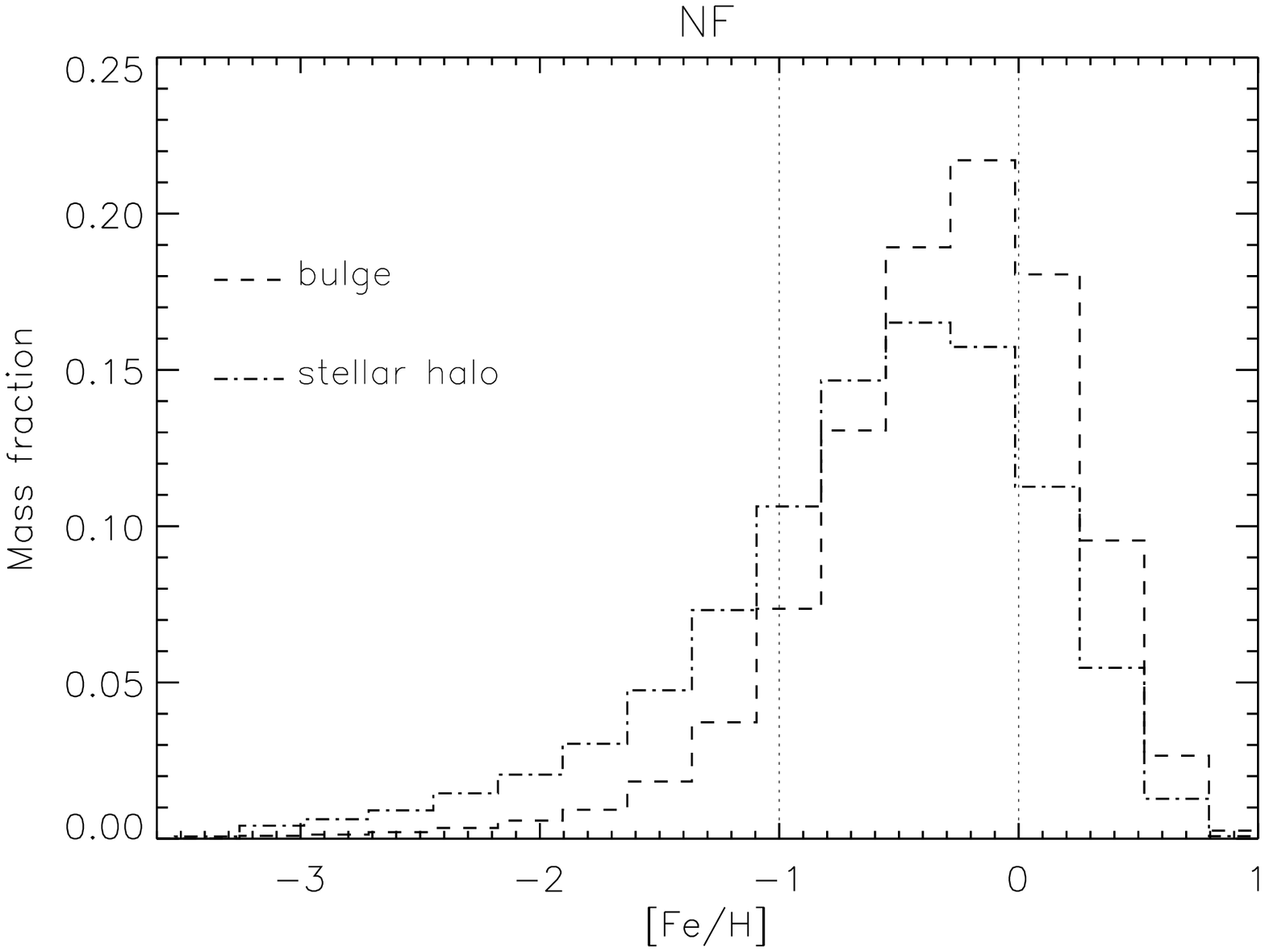}\includegraphics[width=80mm]{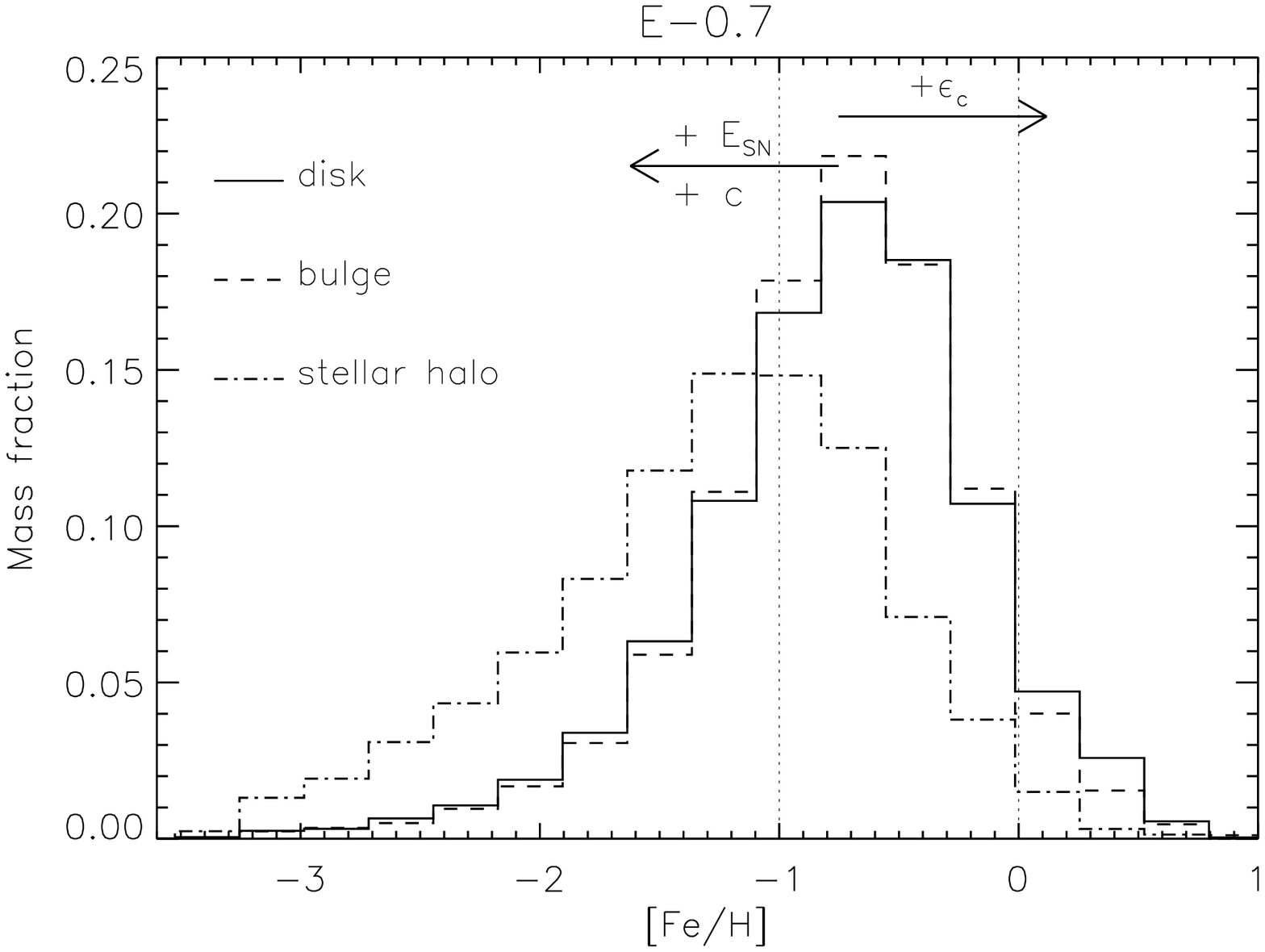}
\caption{Mass fraction as a function of iron abundance for
stars in our no-feedback simulation (NF, left-hand panel), and in
 E-0.7 (right-hand panel panel). We show separately the results for
stars in the disk (for E-0.7), the bulge and the stellar halo.
Vertical lines indicate [Fe/H]=-1 and [Fe/H]=0.}
\label{hist_fe}
\end{figure*}

Most observations of detailed chemical abundances
in stars come from our Galaxy, and are usually quantified
separately for the different dynamical components:
thin disk, thick disk, bulge and stellar halo.
The stellar halo 
is found to be the most metal-poor component, the thin disk and the bulge are
the most metal-rich components  (e.g. Ryan \& Nories 1991; Chiba \& Beers 2000; 
Zoccali et al. 2003; Nordstr\"om et al. 2004).
In order to study the chemical properties of our
simulated galaxies and to see if we can reproduce
these trends,
we use the disk-spheroid decomposition
explained above. 
For simplicity, we do not attempt to
 divide our disk stars into
two components. In the case of the spheroidal component, however,
we have subdivided the stars into  bulge and  stellar halo,
according to distance from the centre, separating at $5\ h^{-1}$ kpc.

In the left-hand panel of Fig.~\ref{hist_fe} we show the distribution
of iron abundance for stars in the  bulge and stellar halo for the
no-feedback case while the right-hand panel shows similar histograms
for E-0.7 together with an abundance histogram for its disk stars.
In both simulations the outer part of the spheroid ("the stellar
halo") is shifted to lower metallicities than the inner part ("the
bulge").
The shift is more pronounced in the simulation with feedback, although 
E-0.7 has a lower level of enrichment overall in comparison to
the no-feedback case. 
The disk and bulge components show
similar iron abundances in E-0.7, while
stars in the  stellar halo have lower abundances, in qualitative
agreement with
 observation.  Note that the metallicities of the inner
stars (i.e. bulge and disk) are the result of  chemical
evolution in the main galaxy, since these components were mainly
formed in-situ. On the contrary, stars in the stellar halo
were mostly contributed by satellites and do not
reflect the chemical evolution of the main component of the galaxy.

The arrows in the right panel  of Fig.~\ref{hist_fe} 
indicate how
the distributions change for different
input parameters. The trends are similar  in all cases.
An increase in $\epsilon_c$ leads
to a shift towards higher iron abundance. 
Note that for higher $\epsilon_c$ more  metals are dumped 
into the cold gas from which stars  form, leading naturally to
 higher stellar  abundances.
On the contrary, increasing the  energy per SN 
decreases the metal abundances because fewer stars
are produced and the gas is less enriched where stars do form.
As shown before, increasing the SF efficiency
leads to lower stellar masses in our model. Consequently,
metallicities are also shifted towards lower values in this case.

For the gas component, the resulting metal distribution
is affected not only by the metal production itself,  but also
by the galactic winds which  transport an important
fraction of the metals outwards, mixing them with the interstellar
gas. For strong enough winds, metals can  be lost into the intergalactic
medium. Hence, many factors influence gas
metallicities.
In Table~\ref{disk_spheroid} we show the fraction of galactic metals
which are locked into the stellar and gas components.
In the no-feedback case, we find that $99$ percent of all
metals are locked into stars. 
This result is because  
i) most (cold) gas in the central regions has been
transformed into stars and hence most metals are locked into
the stellar component, and ii) no winds can develop
in this case so metals cannot be transported outwards.
Consequently, gas in the outer regions (where star
formation is less efficient) is not enriched.
On the contrary, when our feedback model is included,
the fraction of metals locked into stars significantly decreases,
being typically of the order of $20-30$ percent.
Because of the efficiency of SN feedback in blowing
galactic winds and  regulating star formation, most of the metals are now in the
gas component.

\begin{figure}
\includegraphics[width=80mm]{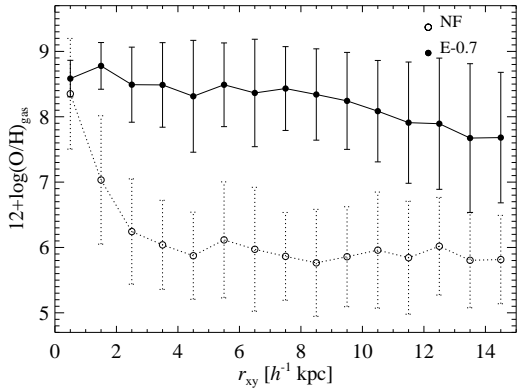}
\caption{Oxygen abundance for the gas component as a function
of radius projected onto the disk plane for our no-feedback simulation
(NF) and for the feedback case E-0.7. The error bars correspond
to the standard deviation around the mean.}
\label{oh_gas_profiles}
\end{figure}

In order to quantify the importance of winds in
 distributing  metals through the
gas component,
the last two columns of Table~\ref{disk_spheroid} 
show the fraction of metals in the gas 
located in the inner regions ($r<2\ r_{\rm opt}$, $f^{\rm
  met}_{\rm gas, in}$)
and in the outskirts of the systems ($2\ r_{\rm opt}<r<r_{200}$, 
$f^{\rm met}_{\rm gas, out}$).  
In the no-feedback case,
only $1$ percent of the 
metals are in the gas. For our feedback cases we find typically  that $70$
percent  of the gaseous metals are outside
the innermost region.
In Fig.~\ref{oh_gas_profiles} we show the oxygen profiles for
the gas components in simulations NF and E-0.7.
SN feedback strongly affects the
chemical distributions. If no-feedback is included,
the gas is enriched only in the very central
regions. Including SN feedback triggers a redistribution
of mass and metals through galactic winds and fountains,
giving the gas component a much higher level of enrichment
out to large radii. A linear fit to this metallicity
profile gives a slope of $-0.048$ dex kpc$^{-1}$ and a zero-point
of $8.77$ dex, consistent with the observed values in real
disk galaxies (e.g. Zaritsky et al. 1994).

We have shown that our star formation and feedback model produces
 galaxies with well differentiated stellar dynamical
components which also differ in their chemical abundances.
In addition, it can produce the correct gas metallicity profiles,
although our chemical model is still very simple.
Galactic winds produced by SN feedback redistribute 
metals  within galactic haloes. 
This is an important success of the model. It opens up the possibility of studying
the detailed distribution of metals within galaxies forming in  their
full cosmological context, as well as exploring the enrichment
of the intergalactic medium.

\section{Dependence on galaxy mass}
\label{dwarf}

\begin{figure*}
\includegraphics[width=80mm]{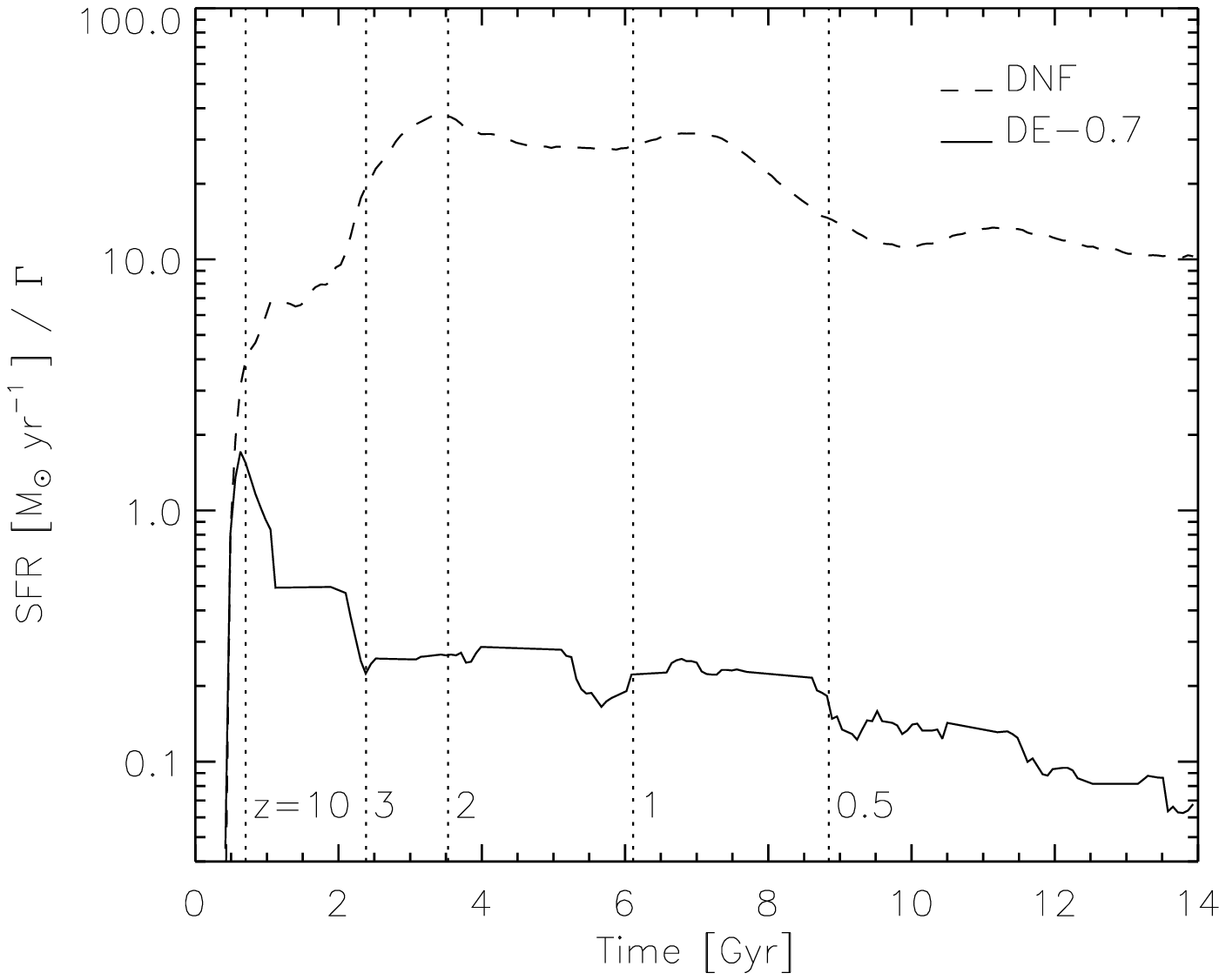}\includegraphics[width=80mm]{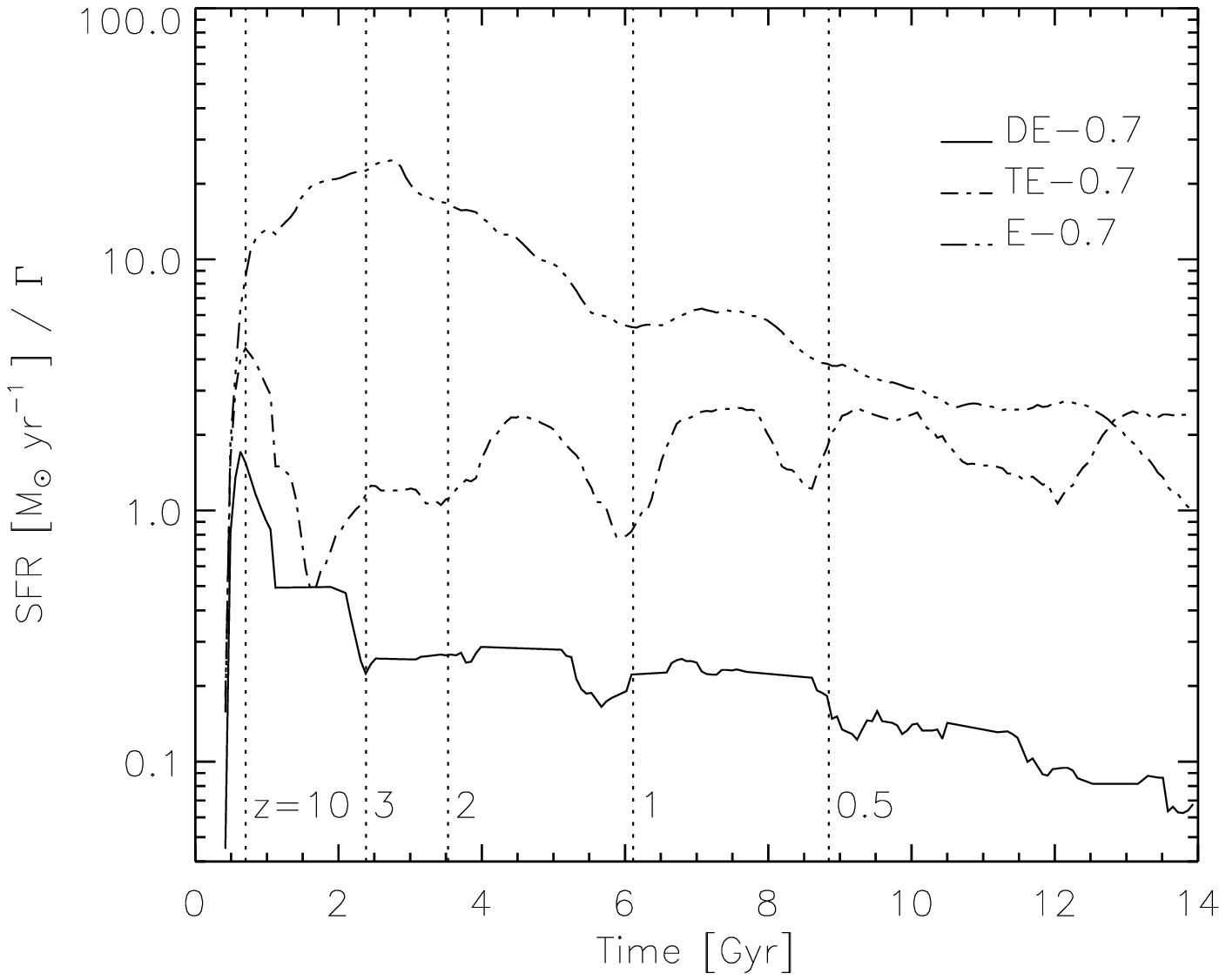}
\caption{SFRs for simulations  of
a  $\sim 10^{9}\ h^{-1}$ M$_\odot$ virial mass system run without
(DNF) and with (DE-0.7) energy feedback,
as well as  for  TE-0.7, a  $\sim 10^{10}\ h^{-1}$ M$_\odot$
simulation including feedback. To facilitate comparison, the SFRs are
normalized to the scale factor $\Gamma$.}
\label{sfr_dwarf}
\end{figure*}

In this Section we investigate how the effects of
SN feedback change with galaxy mass. In Scannapieco
et al. (2006) we showed that the star formation
is affected differently by SN feedback in systems of differing
mass.
We found that for large systems (virial masses
of order  M$_{\rm vir}\sim 10^{12}\ h^{-1}$M$_\odot$),
SN feedback reduces and prolongs star formation with respect
to simulations where it is not included.
In smaller systems (M$_{\rm vir}\le 10^{10}\
h^{-1}$M$_\odot$),
star formation  not only decreases but
its character also changes, becoming much more bursty.
Small systems are very strongly affected
by SN feedback, developing  galactic
winds that are able to expell most of their baryonic
mass.
In this Section, we again test the effects of SN feedback on smaller
mass systems in order to see if the results found
in Scannapieco et al. (2006) are valid for galaxies forming in their
proper cosmological context.

We have scaled down the initial conditions for simulation
E-0.7 by factors of $\Gamma=10^{-3}$ and $10^{-2}$ 
in mass and $\Gamma^{1/3}$ in linear scale to generate
two smaller systems of  final virial masses
$\sim 10^{9}\ h^{-1}$ M$_\odot$  (DE-0.7)
and $\sim 10^{10}\ h^{-1}$ M$_\odot$ (TE-0.7).
These simulations 
were run with the same input physics as E-0.7. 
These initial conditions allow us to
compare simulations of galaxies of  very different mass  but identical
 assembly history.
Note however they may have biased environments and assembly
histories compared to real dwarfs.
We have also run a simulation  of the
$\sim 10^{9}\ h^{-1}$ M$_\odot$  virial mass galaxy
without energy feedback from SN (DNF).
This is  a low mass  counterpart of NF.
In Table~\ref{simulations_dwarf} we indicate the scale factor $\Gamma$, as well
as the dark matter,
gas and stellar masses of the final $z=0$ galaxies renormalised  by the scale factor 
for these  simulations. This is equivalent to expressing masses 
in units of $10^7\ h^{-1}$M$_\odot$ for DNF and DE-0.7,
and in units of  $10^8\ h^{-1}$M$_\odot$ for  TE-0.7.
We also list the gas and baryon fractions for the final galaxies in
these simulations.
All these quantities are computed within the virial radius.
For comparison, in the lower row we repeat the results for  E-0.7
from Table~\ref{simulations_table}.
In this way, the results shown in this table can be
directly compared; any difference reflects
the way SN feedback
varies with  virial mass.

In the left-hand panel of Fig.~\ref{sfr_dwarf}  we show SFRs renormalized by the scale factor 
for the $\sim 10^{9}\ h^{-1}$ M$_\odot$ virial mass
simulations  with and without SN feedback.
SN feedback clearly has a dramatic effect on
the SFR of such  small galaxies.
When SN feedback is not included (DNF), the
SFR behaves similarly to that in  NF (a $10^3$ times more massive galaxy), leading to
a final stellar mass of $1.88\times10^{8}\ h^{-1}$ M$_\odot$.
Within the virial radius, the final mass fraction in stars is
 $0.09$ which is comparable to
that for NF ($0.11$).
The SFR in DE-0.7 behaves quite
differently. It  is very
low at all times; SN feedback  produces  
a decrease in SFR of about 2 orders of magnitude
compared to the no-feedback case.
The final stellar mass formed  is only
$10^{6}\ h^{-1}$ M$_\odot$  ($\sim 1200$ star particles)
in this case.
The amount of gas  within the virial
radius is also small, $M_{\rm gas} = 3.7\times10^7\ h^{-1}$M$_\odot$.
This is the result of the violent winds that develop
in this galaxy due to the
shallow potential well. 
As a consequence, most gas fails to condense and 
form stars.
From the last two columns of Table~\ref{simulations_dwarf},
we can see that the baryons in  DE-0.7 represent only $2$ percent
of the final mass in the galaxy, as opposed
to the case without SN feedback where
the final baryon fraction is $10$ percent.

In the right-hand panel of Fig.~\ref{sfr_dwarf}, we show  SFRs
renormalised by the scale factor for  DE-0.7, TE-0.7 and E-0.7
which differ only in virial mass and were all run with
the same SN feedback model with identical input parameters. 
The effects of SN feedback increase very substantially with decreasing virial mass.
Table~\ref{simulations_dwarf} shows that
more massive galaxies at $z=0$ tend to have 
larger gas and stellar mass fractions ($M/\Gamma$). This is because
more of their gas is able to cool and condense
to build up the stellar component, and less gas
is lost in galactic winds.
For these reasons,  the larger the galaxy, the lower
the final gas fraction (gas over baryonic mass), and 
the larger the baryon
 fraction (baryonic over total mass).

We note that poor resolution may affect our results for the
smallest galaxy. In Scannapieco et al. (2006) we carried out
resolution tests of our feedback implementation.
This analysis suggested that small galaxies must be represented by
many particles (at least $40000$) to get convergent results because
their final state is heavily influenced by details of the interplay
between heating and cooling. Here our galaxies form through both
merging and accretion and it is not clear how much poor resolution
at high redshift affects our results at $z=0$.
In the simulations presented in this paper, all the large galaxies, as
well the smaller object DNF, are represented at $z=0$ by more than
$100000$ dark matter and $70000$ baryonic particles. Only in runs
DE-0.7 and TE-0.7 is the number of baryonic particles ($~17000$ and
$~40000$, respectively) so small that better resolution is probably
needed to obtain full convergence.  However, although the final
properties of the systems may slightly differ at higher resolution, our general
conclusions about the dependence of SN feedback effects on virial mass
should still be valid.

The results obtained in this section prove
that our model is able to reproduce the expected dependence
 of SN feedback on virial mass
without changing the relevant physical parameters.
Thus, our model is well suited for studying the cosmological
growth of structure where large
systems are assembled through mergers of smaller
structures and systems form simultaneously over a wide
range of scales.

\begin{table}
\begin{small}
\caption{List of properties of the simulations with final virial mass   $\sim
10^{9}\ h^{-1}$ M$_\odot$
and  $\sim 10^{10}\ h^{-1}$ M$_\odot$. 
We show  the scale factor $\Gamma$ relative to E-0.7, the dark matter, gas and stellar
masses renormalised by the scale factor $\Gamma$ in units of $10^{10}\ h^{-1}$M$_\odot$.
We also show the gas  
and baryonic fractions within the virial radius.
For comparison, we show the values for E-0.7 
from Table~\ref{simulations_table}.}
\vspace{0.1cm}
\label{simulations_dwarf}
\begin{center}
\begin{tabular}{lcccccc}
\hline
Test  & $\Gamma$ & $M_{\rm DM}^{200}/\Gamma$ & $M_{\rm gas}^{200}/\Gamma$ & $M_{\rm star}^{200}/\Gamma$  & $f_{\rm g}^{200}$ & $f_{\rm bar}^{200}$\\\hline

DNF     & $10^{-3}$ & 179.5   & 2.1 & 18.8 & 0.10 & 0.10\\ 
DE-0.7  & $10^{-3}$ & 176.6   & 3.7 & 0.1  & 0.97 & 0.02\\
TE-0.7  & $10^{-2}$ & 185.2   & 5.2 & 1.6  & 0.76 & 0.04\\
E-0.7   & $1$       & 192.9   & 6.1 & 9.1  & 0.40 & 0.07\\\hline
\end{tabular}
\end{center}
\end{small}
\end{table}

\section{Conclusions}
\label{conclu}

We have used a series of simulations of the formation
of a galaxy of virial mass $\sim 10^{12}\ \ h^{-1}$ M$_\odot$ 
from cosmological initial conditions  to study
how SN feedback affects the formation
of disks. 
Our simulations have sufficient resolution
to resolve the internal structure of the galaxy
as well as  large-scale processes such as mergers, interactions, 
infall, tidal stripping, etc.
Our study has focussed on the performance
of our SN feedback scheme for realistic initial conditions
and on the dependence of the results on the input parameters
of the feedback model, as well as on the mass of the galaxy.
In the following we summarize our main results.

1) Our model for SN feedack is able to regulate
 star formation activity
by heating and pressurizing
gas near star-forming sites. This is reflected in a bursty
behaviour for the SFR when feedback is included, particularly
in low-mass systems. Different
choices for feedback efficiencies produce variations
in the star formation levels and in the detailed shape
of the SFR, although qualitatively all our simulations
show similar behaviour. When compared with a simulation 
where SN feedback is neglected,
the reduction in the final stellar mass formed can be
as high as $70$ percent even for  a Milky Way mass object.

2) Efficient regulation of star formation
leads to an increase in
the final gas fraction, when compared to a simulation
where SN feedback is not included. 
This, together with  substantial mass-loaded galactic
winds powered by the SN input,
leads to galaxies that, at  $z=0$,
have lower  baryon fractions.

3) Galaxy morphology is strongly affected by SN feedback. 
Large disks can be formed and can survive
until $z=0$ if energy injection is  allowed. Our results indicate that 
 disks can, however, be formed for a wide
range  of input parameters.
The stellar mass
in our galaxies is typically dominated by spheroidal components. In
the best case,  we found equal masses in the disk and the spheroid. 
Note, however, that the disk is significantly more luminous
since it is significantly younger. 
We find the
final characteristic size and thickness of disks to depend
significantly on the input
parameters. 
Adopting too extreme 
and unrealistic assumptions for the energy
per SN or for the
efficiency results in  failure to form a disk.
Similarly, a simulation with no-feedback also leads to 
a pure spheroidal stellar component.

4) Disks are rotationally supported, 
have high specific angular momentum and have a substantial number
of young stars. 
Our disks have typical half-mass formation times 
of $\sim 6$ Gyr. In contrast, spheroids
are old, with half-mass formation times $< 3$ Gyr.
In a simulation without feedback much of
the specific angular momentum of the cold component
(the cold gas and stars) 
 is lost to dynamical 
friction during the assembly. As a result,  no disk forms.

5) Our disks  and spheroids have different
chemical properties as a result of their different
formation histories. In particular, 
the stellar halo in our simulations is
the most metal-poor component, in agreement with observation,
 whereas the disk and bulge components
have similar near-solar metallicities. Without SN feedback 
this trend is not recovered and
all components show similar, overly high iron
abundances. Gas metallicities are also
affected by changes in star formation histories
and by galactic winds  when SN feedback is included.
Gas metallicity profiles consistent with observations
are only found when feedback is included.

6) In our model, the effects of SN feedback strongly depend
on virial mass.
We have shown this by scaling down one
of our simulations by factors of $10^2$ and $10^3$ in mass. 
Feedback has  a much stronger effect in smaller
systems, where violent galactic winds
eject most of the baryonic material and
prevent gas from collapsing to make stars.
SFRs are low and bursty in such systems.
Such small galaxies have lower baryon fractions at $z=0$
than larger systems.

Our findings show that our model is able
to produce realistic disk systems from cosmological initial conditions.
This results from
the self-regulation of the star formation
process and the heating due to feedback which prevents excessive early
condensation of gas into the central regions
of dark haloes and the consequent later  loss of angular momentum.
An important success is the fact that
disks are obtained for a variety of
input parameters, with realistic
radial and vertical sizes, and metallicity structures.
In all cases we find that
the spheroids dominate the stellar mass,
suggesting that further refinements are needed if we are to understand
the formation of late-type spirals.

\section*{Acknowledgments}

We thank the referee, Fabio Governato, for useful comments that
helped improve the paper.
We thank Adrian Jenkins and Felix Stoehr for making the initial
conditions available to us.
CS thanks Gabriella De Lucia, Dimitri Gadotti, Klaus Dolag
and Sebasti\'an Nuza for useful and stimulating discussions.
This work was partially supported  by the European Union's ALFA-II
programme, through LENAC, the Latin American European Network for
Astrophysics and Cosmology. Simulations were run 
on Ingeld PC-cluster funded by
Fundaci\'on Antorchas. We acknowledge support from
Consejo Nacional de Investigaciones Cient\'{\i}ficas y T\'ecnicas,
Agencia de Promoci\'on de Ciencia y Tecnolog\'{\i}a and  Fundaci\'on Antorchas.

\end{document}